\pdfoutput=1
\documentclass[twocolumn,nofootinbib,9pt]{revtex4-1}
\usepackage{amsmath,amssymb}    % need for subequations
\usepackage[dvipdfmx]{graphicx}   % need for figures
\usepackage{verbatim}   % useful for program listings

\usepackage{txfonts}
\usepackage{color}
\usepackage{bm}
\usepackage{afterpage}%改ページできるようになる
\usepackage{ascmac}%囲い枠が使えるようになる
\usepackage{listings}%ソースコードが書けるようになる
\usepackage[caption=false]{subfig}

\begin{document}

\title{Characterizing Cryptocurrency market with L\'evy's stable distributions}
\author{Shinji Kakinaka and Ken Umeno}
%\author{Shinji Kakinaka\thanks{kakinaka.shinji.35e@st.kyoto-u.ac.jp} and Ken Umeno\thanks{umeno.ken.8z@kyoto-u.ac.jp}}
\affiliation{Department of Applied Mathematics and Physics, \\
Graduate School of Informatics, Kyoto University.}

\date{\today}

\begin{abstract}
The recent emergence of cryptocurrencies such as Bitcoin and Ethereum has posed possible alternatives to global payments as well as financial assets around the globe, making investors and financial regulators aware of the importance of modeling them correctly.
The L\'evy's stable distribution is one of the attractive distributions that well describes the fat tails and scaling phenomena in economic systems.
In this paper, we show that the behaviors of price fluctuations in emerging cryptocurrency markets can be characterized by a non-Gaussian L\'evy's stable distribution with $\alpha \simeq 1.4$ under certain conditions on time intervals ranging roughly from 30 minutes to 4 hours.
Our arguments are developed under quantitative valuation defined as a distance function using the Parseval's relation in addition to the theoretical background of the General Central Limit Theorem (GCLT).
We also discuss the fit with the L\'evy's stable distribution compared to the fit with other distributions by employing the method based on likelihood ratios.
Our approach can be extended for further analysis of statistical properties and contribute to developing proper applications for financial modeling.
\end{abstract}
\maketitle

\section{Introduction}
Cryptocurrencies have attracted considerable attention across the world as a newly emerging financial asset.
The market has grown explosively since 2009 when Bitcoin was released by Satoshi Nakamoto \cite{Nakamoto2009}, with market capitalization temporarily marking over an astounding 200 billion dollars in 2017.
In order to cool down the boom, financial regulators in countries such as China and Korea imposed strict regulations on cryptocurrency transactions.
Prices turned to decline sharply at the beginning of 2018, and the extreme fluctuations raised the concerns of market volatility.
One feature of this immature asset is the market's decentralized financial system supported by the block-chain technology based on the peer-to-peer network--- different from the central management system seen in central banks.
Another significant feature is that block-chain technology also provides assurances of anonymity, and contributes to a sophisticated system with a well-founded security \cite{Bohme2015}.
Since this alternative system allows for reliable transactions without an intermediary, cryptocurrencies are expected to prevail as an expedient medium of exchange.
Thus, examining price fluctuations of new assets would provide us with some guidance for implementing financial management as well as keys to understand the phenomena occurring in financial systems.

Analysis of financial price fluctuations has long been assumed to follow a Gaussian distribution for its simplicity and the background of the Central Limit Theorem (CLT).
As an example, the famous Black-Scholes model \cite{Blackscholes1973} was formulated under this assumption.
However, it is well known that Gaussianity fails to capture volatile observations and leads to underestimating tail risks.
Extreme fluctuations have been observed repeatedly in financial markets: notable examples include the financial crisis of 2007-2008, which caused turbulence of the market.
Physical (or econophysics) concepts have been offering useful tools for analyzing such economic phenomena.
In the past decades, there have been studies giving an account of asset returns well complied with a L\'evy's stable distribution, which has fatter tails with power-functions compared to a Gaussian distribution \cite{Mandelbrot1963, Fama1965, Hsu1974, MantegnaStanley1995}.
It is one of the most famous parametric fat-tailed distributions and allows us to model not only financial modeling but also a wide range of scientific fields from natural phenomena to computational science \cite{Xu2011, MenabdeSivapalan2000, Koblents2016, Chronis2016, Scalas2007}.
A common motivation in these studies is analyzing extreme values observed in social issues and measuring the liquidity conditions in terms of the parameters of stable laws.
Moreover, in a theoretical context, L\'evy's stable distribution is closely related to an essential theorem--- the Generalized Central Limit Theorem (GCLT) \cite{Gnedenko1954} that thoroughly explains the scaling phenomena in financial markets.
This theorem suggests that the sum of i.i.d. random variables with infinite variance converge only to a L\'evy's stable distribution.
Besides, an extension of the GCLT is studied recently \cite{Shintani2018} with the application of the form of L\'evy's stable distribution.
Such arguments enable us to capture the inherent characteristics of asset price fluctuations and help identify the probability distribution of asset returns.
Thus, analysis of price fluctuation behaviors using L\'evy's stable distribution can be crucial to understand the mechanism of financial markets \cite{Frank2016}.

A paper by Begu\v{s}i\'c et al. \cite{Begusic2018} studies the fat-tailed nature of price fluctuations for Bitcoin and reveals that $\alpha$ is $2.0 \sim 2.5$, by using the traditional Hill estimator method.
The method focuses on finding a local fit for tails, and although the results provide interesting findings of power-law behaviors, it does not account for the entire distribution.
On the other hand, the framework of L\'evy's stable distribution covers the entire dataset, allowing us to investigate extreme and non-extreme price fluctuations from the same standpoint.

In this paper, we analyze the price fluctuation behaviors of emerging cryptocurrency markets with the L\'evy's stable distribution and examine the validity of the model.
We first show that the probability density of price returns are in a good agreement with the L\'evy's stable distribution through the parameter estimation in the case of a fixed 1-hour time interval.
We next consider different time intervals for extensive analyses and provide empirical evidence that price fluctuations in cryptocurrency markets do not follow a Gaussian distribution and can be better described by a L\'evy's stable distribution.
To confirm this, we propose a numerical assessment by using a function representing the distance between theoretical and empirical distributions, which is obtained from the Parseval's relation.
An advantage of this approach is to evaluate stable distributions {\it quantitatively}, and at the same time, to avoid the analytical difficulties.
In addition, we examine the scaling property of returns to check whether the L\'evy stable regime holds.
The combination of these approaches helps lead to a practical analysis for detecting stable laws in cryptocurrency markets.
We discuss that if we admit some intrinsic noise errors, returns can be assumed to follow a L\'evy stable regime within a certain range of time intervals--- outside the range, there are either quantitative or theoretical failures.
Furthermore, we discuss whether the L\'evy's stable distribution can be an appropriate model by examining the model-fitting for returns under the L\'evy's stable distribution and under other fat-tailed distributions.
Our study compares fitting approaches covering the large portion of the distribution with those covering only the tail parts of the distribution, including the Hill estimator.
The idea proposed in this paper is helpful not only to value the liquidity conditions of the market but also provide clues towards financial modeling in a more careful manner.
%%%%%%%%%%%%%%%%%%%%%%%%%%%%%%%%%%%%%%%%%%%%%%%%%%%%
% 2. Methodology
%%%%%%%%%%%%%%%%%%%%%%%%%%%%%%%%%%%%%%%%%%%%%%%%%%%%
\section{Methodology}
This section explains the methods used for analysis in this study.
First, we summarize the basic properties of L\'evy's stable distribution in the first subsection.
In the next subsection we discuss what method applies to parameter estimation.
We report that the method with the use of characteristic function is preferred to the traditional Hill estimator method.
The third subsection introduces a quantitative valuation by means of characteristic function, which can be expected as a tool to evaluate the fit with L\'evy's stable distributions.
Finally, the last subsection describes a method used for testing the fit compared to other forms of distributions.
%%%%%%%%%%%%%%%%%%%%%%%%%%%%%%%%%%%%%%%%%%%%%%%%%%%%
% 2.1 L\'evy's Stable Distribution
%%%%%%%%%%%%%%%%%%%%%%%%%%%%%%%%%%%%%%%%%%%%%%%%%%%%
\subsection{L\'evy's stable distribution}
A L\'evy's stable distribution was first introduced by Paul L\'evy \cite{PaulLevy1937}, with tails that are expressed as power-functions.
It is also called $\alpha$-stable distribution, or stable distribution.
With the constants $c_+ > 0, $ $c_- > 0, $ $\alpha > 0$, the far tails of the probability density function (PDF) can be approximately written as
\begin{align*}
	f(x) \simeq
	\begin{cases}
	c_+ |x|^{-(1+\alpha)} \;\; \mbox{for} \;\; (x \rightarrow +\infty) \\
	c_- |x|^{-(1+\alpha)} \;\; \mbox{for} \;\; (x \rightarrow -\infty).
	\end{cases}
\end{align*}
Stable distributions are defined as the following:
A random variable $X$ is said to be stable and have a stable distribution if there is a positive constant number $c$ and a real number ${d \in \mathbb{R}}$ such that
\begin{align*}
	aX_1+bX_2 \overset{\mathrm{d}}{=} cX+d,
\end{align*}
for positive constant numbers ${a,b}$ and when variables $X_1, X_2$ are i.i.d. copies of $X$.
Here, ${\overset{\mathrm{d}}{=}}$ denotes equality in distribution \cite{SamoTaqqu1994}.
Stable distribution is represented by 4 parameters; $\alpha \in (0,1], \beta \in [-1,1], \gamma >0$, and $\delta \in (-\infty, \infty)$.
When the variable $X$ follows the stable distribution, the notation
\begin{align*}
	X \overset{\mathrm{d}}{=} S(\alpha, \beta, \gamma, \delta)
\end{align*}
is often used.
Here $\alpha$ is the tail index parameter, which indicates the fatness of the tail, $\beta$ the skewness parameter, $\gamma$ the scale parameter, and $\delta$ the location parameter.
Stable distribution has a property that the mean does not exist for ${0 < \alpha \leq 1}$, and the variance diverges for ${0 < \alpha < 2}$.
Furthermore, the PDF cannot be written analytically except for a few cases (${\alpha=2, \beta=0}$; Gaussian distribution, ${\alpha=1, \beta=0}$; Cauchy distribution, ${\alpha=1/2, \beta=1}$; L\'evy distribution).
Instead, it is expressed by the characteristic function ${\phi(k)}$ (CF).
This CF is a Fourier Transformation (FT) of the PDF:
\begin{align*}
	f(x)=\frac{1}{2\pi} \int_{-\infty}^{\infty} \mathrm{e}^{-ikx} \phi(k) dk.
\end{align*}
When the variable $X$ follows ${S(\alpha,\beta,\gamma,\delta)}$, the CF can be shown as
\begin{align}
	\label{eq:first}
	\nonumber
	\phi(k) &=
	\exp \left \{ i\delta k -\gamma^\alpha |k|^\alpha (1-i \beta \operatorname{sgn}(k) \omega(k, \alpha)) \right \}, \\
	\omega(k, \alpha) &=
	\begin{cases}
	\tan( \frac{\pi \alpha}{2} ) & \alpha \neq 1 \\
	-\frac{2}{\pi} \log|k| & \alpha = 1.
	\end{cases}
\end{align}
This is equivalent to the one-parameterization form $S(\alpha,\beta,\gamma,\delta;1)$ for Nolan\cite{Nolan2018}, which is the most common form and is preferred to use when one is interested in the basic properties of the characteristic function.
Note that the distribution is symmetric if $\beta=0$, right-tailed if positive, and left-tailed if negative.
%%%%%%%%%%%%%%%%%%%%%%%%%%%%%%%%%%%%%%%%%%%%%%%%%%%%
% 2.2 Parameter Estimation
%%%%%%%%%%%%%%%%%%%%%%%%%%%%%%%%%%%%%%%%%%%%%%%%%%%%
\subsection{Parameter estimation}
Numerous approaches are known for parameter estimation.
Since the PDF is not always expressed in a closed form, there are some challenges to overcome the analytic difficulties.
This has long been a motivation for researchers to construct a variation of estimation methods, and the representatives are for instance; 
the approximate maximum likelihood estimation \cite{DuMouchel1973, BrorsenYang1990, Mittnik1999, Nolan2001}, non-parametric quantile (QM) method \cite{FamaRoll1971, McCulloch1986}, fractional lower order moment (FLOM) method \cite{Ma1995}, method of log-cumulant \cite{Nicolas2002, Pastor2016}, the characteristic function (CF) based method \cite{Koutrouvelis1980, Bibalan2017, Press1972, Fukunaga2018} and more.

While these methods aim to get estimators related to the stable distribution, there are some methods that can be applied to the case where the data is expected to follow a power-law.
One common approach is the traditional Hill estimator \cite{Hill1975}, which focuses on estimating the tail index parameter $\alpha$.
The approach pays attention to discover the power law decay of the \textit{tail} portion of the cumulative distribution, $P(X>x)\sim x^{-\alpha}$ (then the PDF decays with $\alpha+1$).
This method is known to be a right choice of tool for identifying and qualifying the tail properties in empirical studies \cite{Plerou2008, Gop1999, Begusic2018}, and often reveals the \textit{inverse cubic law} in many financial asset returns.
Before the estimation, one first needs to set the lower bound $x_{\mathrm{min}}$, which means that the power law is studied only for values larger than the lower bound.
The idea is to estimate local slopes of the tail portion of the distribution as,
\begin{align*}
	\hat{\alpha}=n\left(\sum_{i=1}^{n} \ln \frac{x_{i}}{x_{\min }}\right)^{-1},
\end{align*}
where $x_i \ (i=1,2,\ldots,n)$ is the $n$ largest data out of $N$ observed data, such that $x_i \geq x_{\mathrm{min}}$.
Note that the method is based on the technique of {\it maximum likelihood estimator}.
Hill estimator is known to be asymptotically normal and consistent for $n, N\rightarrow \infty, n/N \rightarrow 0$, and the standard error on $\hat{\alpha}$ is $\sigma = \hat{\alpha}/\sqrt{n}$.

The choice of the lower bound $x_{\mathrm{min}}$ is a crucial issue when applying to empirical data.
If we choose $x_{\mathrm{min}}$ too large, estimation for local slopes for the tail portion will be more inaccurate \cite{Clauset2009}.
Fitting local tails becomes difficult because of including other portions of the distribution, which usually tends to show properties different from the tail.
On the other hand, if $x_{\mathrm{min}}$ too small is chosen, we will get a biased estimate due to the lack of sample numbers.
Moreover, the estimator gives excellent results when the data follows a power-law form, but also give some estimation for data that is not necessarily drawn from a power-law distribution.
In other words, the estimator calculates $\alpha$ accurately that best fits the simple power-law form $x^{-\alpha}$for any data in the range of $x \geq x_{\mathrm{min}}$.
Although the far tails of cumulative distribution for stable distribution show the simple power law form as well: $P(X>x)\sim cx^{-\alpha}$, with the constants $c=\Gamma(\alpha)(\sin(\pi \alpha/2))(1+\beta)/\pi$, it tends to have overestimated $\hat{\alpha}$ when choosing the proper $x_{\mathrm{min}}$ is not taken into account\cite{Weron2001}.

To mitigate this issue, we employ the method of estimating the best choice of $x_{\mathrm{min}}$\cite{Clauset2009}, which helps to see whether the Hill estimator is valid for stable distributions.
The idea is the use of the Kolmogorov-Smirnov (KS) statistic, which represents the maximum distance between two distributions in terms of cumulative distribution function (CDF) shown as:
\begin{align*}
	D = \max_{x\geq x_{\mathrm{min}}}|P(x)-Q(x)|,
\end{align*}
where $P(x)$ is the CDF obtained from empirical data, and $Q(x)$ is the CDF that best fits the power law model.
With a given lower bound $x_{\mathrm{min}}$, KS statistic can be obtained using data points in the range of $x\geq x_{\mathrm{min}}$.
The estimation for the lower bound $\hat{x}_{\mathrm{min}}$ is then the one that minimizes the KS statistic $D$.
This method gives good results and achieves to estimate $\hat{x}_{\mathrm{min}}$ precisely and adequately.
However, when the distribution follows a power law only in the limit of very large $x$, it can be unrealistic assuming to fit with distribution $x^{-\alpha}$ for any specific range of $x$.
Stable distributions have forms to illustrate this case; the far tails are equivalent to the pure power-law form $x^{-\alpha}$ but only in an asymptotic behavior.
Therefore, finding the actual value of $x_{\mathrm{min}}$ is obscured by the fact that stable distribution does not exactly correspond to the pure form $x^{-\alpha}$ within ranges of observed values.
Table~\ref{tb:hill} shows estimation results of $\alpha$ for stable distributions based on the method of Hill estimator with the use of KS statistics.
\begin{table}
 \begin{center}
 \caption{
 The average value of 1000 simulated estimates of tail index $\alpha$ for L\'evy stable samples conducted with different sizes of datasets $N$. We also report the range of estimated $\alpha$ with a confidence level of 95\% obtained from simulation to give a view of estimation errors.
 }  
  \begin{tabular}{l p{0mm} c p{0mm} c p{0mm} c} \hline
     true $\alpha$ & & $N=10^3$ & & $N=10^4$ & & $N=10^5$\\ \hline \hline
     & & \multicolumn{5}{c}{average value of $\hat{\alpha}$ and its 95\% confidence intervals}\\ \cline{3-7}
     1.2 & & $1.20$ & & $1.22$ & & $1.21$\\
      & & $(0.98-1.54)$ & & $(1.00-1.45)$ & & $(0.93-1.56)$\\
     1.4 & & $1.50$ & & $1.47$ & & $1.41$\\
      & & $(1.20-1.92)$ & & $(1.18-1.73)$ & & $(1.11-1.77)$\\
     1.6 & & $1.94$ & & $1.81$ & & $1.63$\\
      & & $(1.49-2.60)$ & & $(1.40-2.13)$ & & $(1.31-2.00)$\\
     1.8 & & $2.60$ & & $2.67$ & & $1.87$\\
      & & $(1.86-3.73)$ & & $(1.73-3.10)$ & & $(1.55-2.20)$\\ \hline
      & & \multicolumn{5}{c}{95\% confidence intervals of $n$ for the lower bound $\hat{x}_{min}$}\\ \cline{3-7}
     1.2 & & $(57-267)$ & & $(64-256)$ & & $(28-136)$ \\
     1.4 & & $(45-245)$ & & $(80-344)$ & & $(40-170)$ \\
     1.6 & & $(37-218)$ & & $(63-459)$ & & $(52-214)$ \\
     1.8 & & $(25-195)$ & & $(54-603)$ & & $(69-285)$ \\
     \hline
  \end{tabular}
  \label{tb:hill}
  \end{center}
\end{table}
The results indicate that even though we have chosen the lower bound properly that suits well with the power laws, the method still overestimates $\alpha$ for random stable variables $S(\alpha,\beta=0,\gamma=1,\delta=0)$, especially for $\alpha$ larger than 1.4.
As the parameter becomes close to 2, it fails to give a consistent estimator.
What is more, the number of data $n$ used for the estimation of $\alpha$ does not depend on the sample size.
No specific range of $x$ other than for about $50 \sim 500$ extreme values can be more appropriate for detecting the power law for stable distributions.
Although the combination of Hill estimator and the KS statistics can sometimes successfully lead to good estimation results, susceptibilities to estimation errors remain quite large since $n$ can be no larger than around $500$, regardless of the size of datasets.
Therefore, such a technique may present unreliable results under the assumption of stable distributions.

In response to this result, a method that can take enough data into consideration is preferred when dealing with stable distributions.
Many of the representative methods suggested at the beginning of this subsection tend to have several issues, such as a limited range of estimation, a high computational cost, and the requirement of a larger dataset.
The CF-based method makes good use of CF's distinctive features and is most frequently applied for its relatively less defect compared to other methods \cite{Kateregga2017}.
In particular, the regression-based method \cite{Koutrouvelis1980, Fukunaga2018} provides a straightforward approach with the application of regressions using the CF form, which is the estimator of our choice.
It shows fast and accurate computation well enough to estimate cryptocurrency data (For more information about the estimation method, look Appendix A).
%%%%%%%%%%%%%%%%%%%%%%%%%%%%%%%%%%%%%%%%%%%%%%%%%%%%%
% 2.3 Appraisal for the L\'evy stable regime through the characteristic function
%%%%%%%%%%%%%%%%%%%%%%%%%%%%%%%%%%%%%%%%%%%%%%%%%%%%%
\subsection{Appraisal for the L\'evy stable regime through the characteristic function}
For the goodness-of-fit, statistical tests have analytical difficulties in practice due to the lack of fundamental statistics, especially the lack of a closed-form of PDF.
Numerically accurate expressions are known for stable distributions, but often have symmetric constraints \cite{Crisanto2018, Karina2018}.
Therefore, statistical indicators such as KS statistics and KL divergence have fundamental problems to be applied when modeling with stable distributions.
As an alternative, we focus on the CF, following the fact that the inversion formula for the CF indicates a one-to-one correspondence between the PDF and the CF.
The CF of the stable distribution can be expressed analytically as equation~\eqref{eq:first}.
Our attempt here is to calculate the difference or the distance between the PDF of the estimated stable distribution (theoretical) and the PDF obtained from a large number of real data (empirical).
The distance we consider is a simple form shown as,
\begin{align*}
	\int_{-\infty}^\infty \left |\hat{p}(x)-p(x) \right |^2 dx.
\end{align*}
where $\hat{p}(x)$ is the PDF for the estimated stable distribution in a continuous form and $p(x)$ for the empirical distribution as well.
When we discuss the empirical PDF $p_N(x)$ from $N$ observed data, we should consider it in a discrete version due to its discontinuous form.
In the belief that continuous-time signals of the empirical PDF could be discretized into discrete-time signals, we obtain
\begin{align*}
	\sum_{n=-\infty}^{\infty} \left |\hat{p}[n]-p_N[n] \right |^2,
\end{align*}
where ${\hat{p}[n]}$ and ${p_N[n]}$ represents the discretized form of ${\hat{p}(x)}$ and ${p_N(x)}$, respectively.
We do not conduct the process of discretization in practice but instantly use the Parseval's theorem based on Discrete-time Fourier Transform (DTFT), which yields
\begin{align}
	\label{eq:second}
	\nonumber
	\sum_{n=-\infty}^{\infty} \left |\hat{p}[n]-p_N[n]
	\right |^2 &= \frac{1}{2\pi} \int_{-\pi}^{\pi} \left | \hat{\phi}(k)-\phi_N(k)
	\right |^2 dk \\
	\nonumber
	&= \lim_{\Delta k \rightarrow 0} \left ( \frac{1}{2\pi} \sum_{k=-\pi}^{\pi} \left | \hat{\phi}(k)-\phi_N(k)
	\right |^2 \Delta k \right ) \\
	& \simeq \frac{\Delta k}{2\pi} \sum_{i=1}^{\frac{2\pi}{\Delta k}} \left |\hat{\phi}(k_i)-\phi_N(k_i)
	\right |^2.
\end{align}
where $\Delta k$ is the width of bin for Riemann sums.
Note that the Parseval's theorem holds under the assumption of the sampling theorem, which requires sampling intervals to be refined enough.
The process of Riemann sum in equation~\eqref{eq:second} approximately holds when both conditions satisfy: a large enough number of data to obtain an unbiased estimate of $\hat{\phi}(k_i)$, and a small enough width of bin $\Delta k$.
In this paper, for all cases, $\Delta k$ is assumed as small an amount as $2\pi /100$ for computation convenience in the process of summation, which means the distance is calculated as 100 sums of $\frac{1}{100} \left |\hat{\phi}(k_i)-\phi_N(k_i) \right |^2$ for the range of $k_i\in[-\pi, \pi]$ ($k_1 = -\pi, k_2 = -\pi+\Delta k,\ldots, k_{100} = \pi$).
This method implies that the distance between the theoretical and the empirical PDF could be calculated based on the same idea in the form of CF.

Similar function forms are introduced as the minimum distance method for parameter estimation \cite{Press1972, Paulson1975}; however, they have put a weight function to the distance function.
Heathcote extended to develop a more general setting, but the method still has the difficulties of selecting the proper values \cite{Heathcote1977}.
The distance function we propose is advantageous for many application due to its simple form and presents less computational drawbacks.
\begin{figure}[htbp]
  \includegraphics[width=\linewidth]{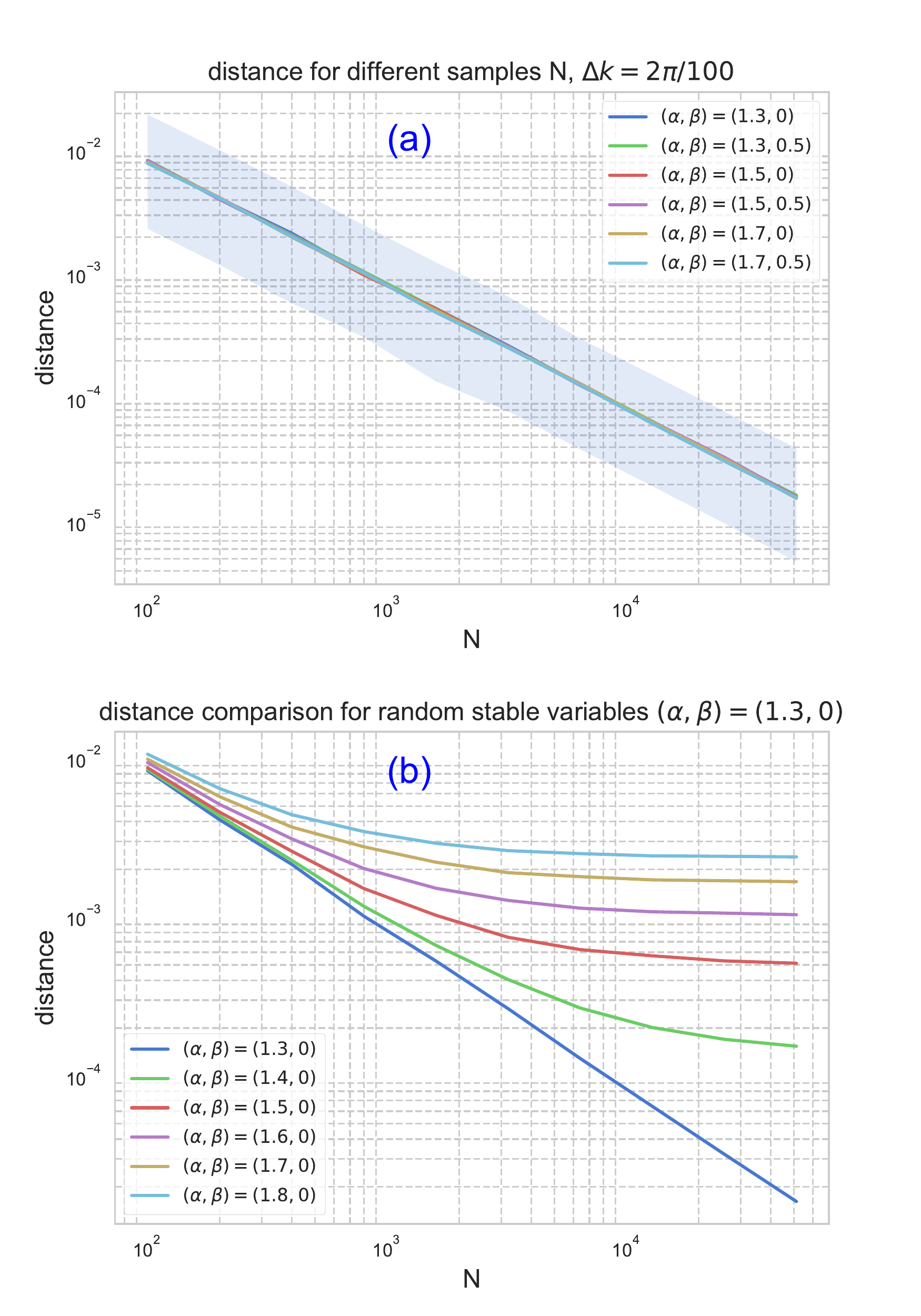}
  \caption{
   (a) represents calculated distance between the theoretical stable distribution and the empirical distribution derived from the original stable distribution.
   First, a number of $N$ synthetic data is generated by the stable random generator method (Appendix B).
   Then the distance is calculated as shown in equation~\eqref{eq:second}.
   Theoretical values for the theoretical stable distribution are given for different combinations of the parameters $(\alpha, \beta)$.
   For the effects of bias, we show the average of 1000 simulated distance associated with the 95\% confidence intervals of the synthetic 1000 distances (shadowed in light blue).
   Simulation results show that distance depends on the number of data $N$, and the average decreases with the order $\mathcal{O}(1/N)$, as demonstrated in equation~\eqref{eq:third}, with deviation error that also decreases with the order $\mathcal{O}(1/N)$.
   (b) shows the behavior of the distance between the average of 1000 generated stable distributions and the theoretical stable distribution fixed to $S(1.3, 0, 1, 0)$, as the number of data $N$ changes.
   The calculated average distance converges to the actual value for a more significant number of datasets.
   Most notably, the more $\alpha$ drifts away from 1.3, the larger the distance becomes for any number of data.
   }
   \label{distance1}
\end{figure}

Next, we remark on the validity of the distance function.
We check the applicability of the definition of the distance function to make further discussions possible for fitting data to stable laws.
Figure~\ref{distance1} shows the basic properties and results needed for explaining the concept.
Sub-figure (a) shows the simulated distance between theoretical stable distributions and generated stable distributions.
The deviation error for distance (bias from finite-size effects) is also shown.
Here, the random generator for stable distributions is based on the method proposed by Weron \cite{Weron1996} (Appendix B).
When $\phi_N(k)$ obtained from i.i.d. distributed data $X_n$ ideally follows some theoretical distribution and becomes the true value $\phi(k)$ as $N\rightarrow \infty$, it can be shown that 
\begin{align}
\label{eq:third}
    %E\left[\left|\phi_{N}(k)\right|^{2}\right]=|\phi(k)|^{2}+\frac{1}{N}\left(1-|\phi(k)|^{2}\right)
	E\left[\left|\phi(k) - \phi_{N}(k)\right|^{2}\right]=\frac{1}{N}\left(1-|\phi(k)|^{2}\right),
\end{align}
where $E[\cdot]$ is the expectation with respect to the data distribution.
Then the expectation of the distance function is the average of~\eqref{eq:third} for $k\in [-\pi, \pi]$, which decreases with the order $\mathcal{O}\left (1/N \right )$ (Appendix C).
The bias of distance decreases with the same order $\mathcal{O}\left (1/N \right)$, which is clarified through simulation.
Sub-figure (b) checks if there is no inconsistency between the theoretical distance and the synthetic distance.
According to the simulation results, we know that distance function is generally independent of the parameters $(\alpha, \beta)$, as well as showing larger values for a stronger degree of distribution transformers.
No exceptional or inconsistent results are observed, which indicates that it can potentially be used as an appropriate tool to obtain a numerical expression in order to grasp the relations of distributions.
When a sufficient amount of data is given or known, the distance 
is determined and can be obtained as a particular value.
This value not only indicates simply the distance between the two distributions but also be a standard measure to discuss error evaluations of the calculated distance.
%%%%%%%%%%%%%%%%%%%%%%%%%%%%%%%%%%%%%%%%%%%%%%%%%%%%%
% 2.4 Comparison with alternative distributions
%%%%%%%%%%%%%%%%%%%%%%%%%%%%%%%%%%%%%%%%%%%%%%%%%%%%%
\newpage
\subsection{Comparison with alternative distributions}
Although the parameter estimation method and our evaluation method proposed in the previous subsection illustrate how to analyze data with stable distributions, they may still be unsatisfactory for discussing the validity of the model.
These methods find and evaluate the best fit under the condition of stable laws, but it does not necessarily mean that the stable distribution exactly describes the data.
Thus, we compare the model-fits under the stable distribution with those under other distributions.

Regardless of how well the empirical data fit with a stable distribution, the data may fit more with other distributions.
An alternative distribution, for instance, a power-law or exponential distribution, may show a better fit.
Even when the data does not follow any typical form of distribution, or when the exact distribution cannot be identified empirically, this approach tells us which model can be reasonable for the fit.
Here we employ a {\it likelihood ratio test}, applied by Clauset et al.,~\cite{Clauset2009} to directly compare two candidate distributions against each other and decide which distribution provides a better fit.
The method is based on calculating the likelihood of the data.
With given PDFs of $p_1(x)$ and $p_2(x)$, the log-likelihood ratio is obtained as
\begin{align}
\label{eq:likelihood}
\mathcal{R} = \ln \prod_{i=1}^{n} \frac{p_{1}\left(x_{i}\right)}{p_{2}\left(x_{i}\right)} = \sum_{i=1}^{n}\left\{\ln p_{1}\left(x_{i}\right)-\ln p_{2}\left(x_{i}\right)\right\},
\end{align}
which is equivalent to the logarithm of the ratio of the two likelihoods.
As a higher likelihood indicates a better fit, a positive value of $\mathcal{R}$ implies that the former distribution is better than the latter.
Thus, the ratio value of $\mathcal{R}$ can be an indicator for judging which distribution is efficient for the fit.

In practice, making a judgment is difficult when $\mathcal{R}$ is close to zero, almost in the event of a tie, since the results depend on statistical fluctuations of the likelihood values.
To avoid misclassification, we calculate the $p$-value, associated with the normalized log-likelihood ratio $\mathcal{R}/ \left(\sigma \sqrt n \right)$ ($\sigma$ is shown in equation~\eqref{eq:sigmaeq}), to confirm whether the obtained ratio shows a statistically significant result (see Clauset et al.~\cite{Clauset2009} for more).
The $p$-value can be calculated as the probability that the log-likelihood ratio becomes larger than the absolute value of observed $\mathcal{R}$.
The sum of i.i.d. observations, $\mathcal{R}$, becomes normally distributed by the CLT.
Thus the value is calculated as
\begin{align}
	\label{eq:pvalue}
	p =\operatorname{erfc} \left(|\mathcal{R}| / \sqrt{2 n} \sigma \right),
\end{align}
where $\operatorname{erfc}$ denotes the complementary Gaussian error function,
\begin{align*}
	\operatorname{erfc}(z) = \frac{2}{\sqrt{\pi}} \int_{z}^{\infty} \mathrm{e}^{-t^{2}}dt,
\end{align*}
and $\sigma$ denotes the estimated standard deviation of a single term on $\mathcal{R}$:
\begin{align}
	\label{eq:sigmaeq}
	\sigma^{2}=\frac{1}{n} \sum_{i=1}^{n}\left\{\left(\ln p_1(x_{i})-\ln p_2(x_{i})\right)-\left(\overline{\ln p_1(x)}-\overline{\ln p_2(x)}\right)\right\}^{2},
\end{align}
where bar denotes the average of terms.
If the value is small enough ($p<0.1$), the result is statistically significant.
In this case, it is sufficient to make a judgment for discriminating which distribution model is proper for fitting the data.
%%%%%%%%%%%%%%%%%%%%%%%%%%%%%%%%%%%%%%%%%%%%%
% 3. Analysis of Cryptocurrency
%%%%%%%%%%%%%%%%%%%%%%%%%%%%%%%%%%%%%%%%%%%%%
\section{Analysis of Cryptocurrency}
In this section, four subsections are beginning with the presentation of 5 types of cryptocurrency datasets for analyzing returns.
The second subsection shows the results of the parameter estimation for cryptocurrency returns with a time scale of $\Delta t = 1 \mathrm{hour}$ when explained by a stable distribution.
Furthermore, returns for different time scales are discussed in the third subsection in terms of the estimated index parameter $\alpha$ and the distance measure.
We strengthen the importance of time scaling for the stable model but also address the issues for practical use and applications.
The last subsection shows the comparison of the model with other representative fat-tail models to discuss the validity of the stable distribution for cryptocurrency returns.
%%%%%%%%%%%%%%%%%%%%%%%%%%%%%%%%%%%%%%%%%%%%%%%%%%%%%
% 3.1 Cryptocurrency data presentation
%%%%%%%%%%%%%%%%%%%%%%%%%%%%%%%%%%%%%%%%%%%%%%%%%%%%%
\subsection{Cryptocurrency data presentation}
This subsection explains the basic characteristics of our data on cryptocurrencies.  
Table \ref{tb:jikasougaku} shows the market capitalization and the price of 5 major cryptocurrencies, Bitcoin (BTC); Ethereum (ETH); Ripple (XRP); Litecoin (LTC) and Monero (XMR).
Data is taken from Cryptocurrency Market Capitalization (https://coinmarketcap.com).
\begin{table}
 \begin{center}
 \caption{Basic data facts of cryptocurrencies (2019/01/15)}
  \begin{tabular}{l p{5mm} r p{5mm} r} \hline
     Cryptocurrency & & Market Cap[\$] & & Price[\$]\\ \hline \hline
     Bitcoin (BTC) & & $64,308,311,082$ & & $3,678.28$\\
     Ethereum (ETH) & & $13,391,497,879$ & & $128.29$\\
     Ripple (XRP) & & $13,534,746,905$ & & $0.33$\\
     Litecoin (LTC) & & $1,938,420,144$ & & $32.29$\\
     Monero (XMR) & & $761,083,680$ & & $45.58$\\
     \hline
  \end{tabular}
  \label{tb:jikasougaku}
  \end{center}
\end{table}

Bitcoin is the most dominant cryptocurrency, whereas the others are considered as minor coins.
However, recently, some minor coins (alto-coins) such as Ripple and Litecoin have also emerged rapidly since the arrival of the cryptocurrency boom in the mid 2017.
They have attracted considerable attention; market capitalization reached a peek more than a billion dollars momentarily.
Given the impacts of the cryptocurrency market on the economy, the importance of analyzing alto-coins has increased greatly.
%%%%%%%%%%%%%%%%%%%%%%%%%%%%%%%%%%%%%%%%%%%%%
% 3.2 Results
%%%%%%%%%%%%%%%%%%%%%%%%%%%%%%%%%%%%%%%%%%%%%
\subsection{Parameter estimation results}
From the data above, we estimate the parameters of the stable distribution that best describes the empirical returns.

\begin{figure}
 \subfloat[][BTC ($\alpha$=1.327, $\beta$=$-$0.028), $\Delta t = 1$ hour]{
  \includegraphics[width=90mm,bb=9 9 560 280]{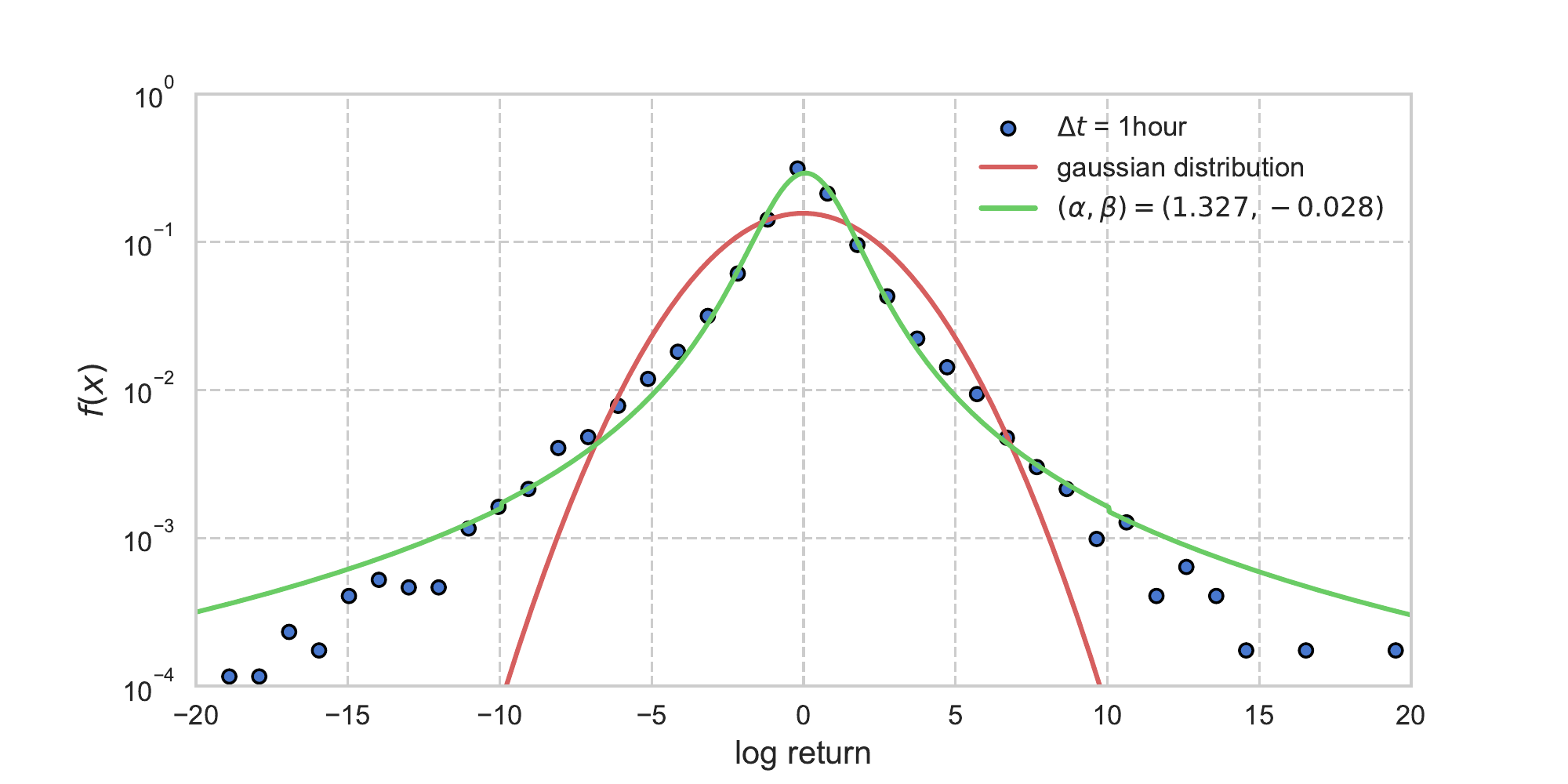}} \\
\subfloat[][LTC  ($\alpha$=1.411, $\beta$=0.018), $\Delta t = 1$ hour]{
  \includegraphics[width=90mm,bb=9 9 560 280]{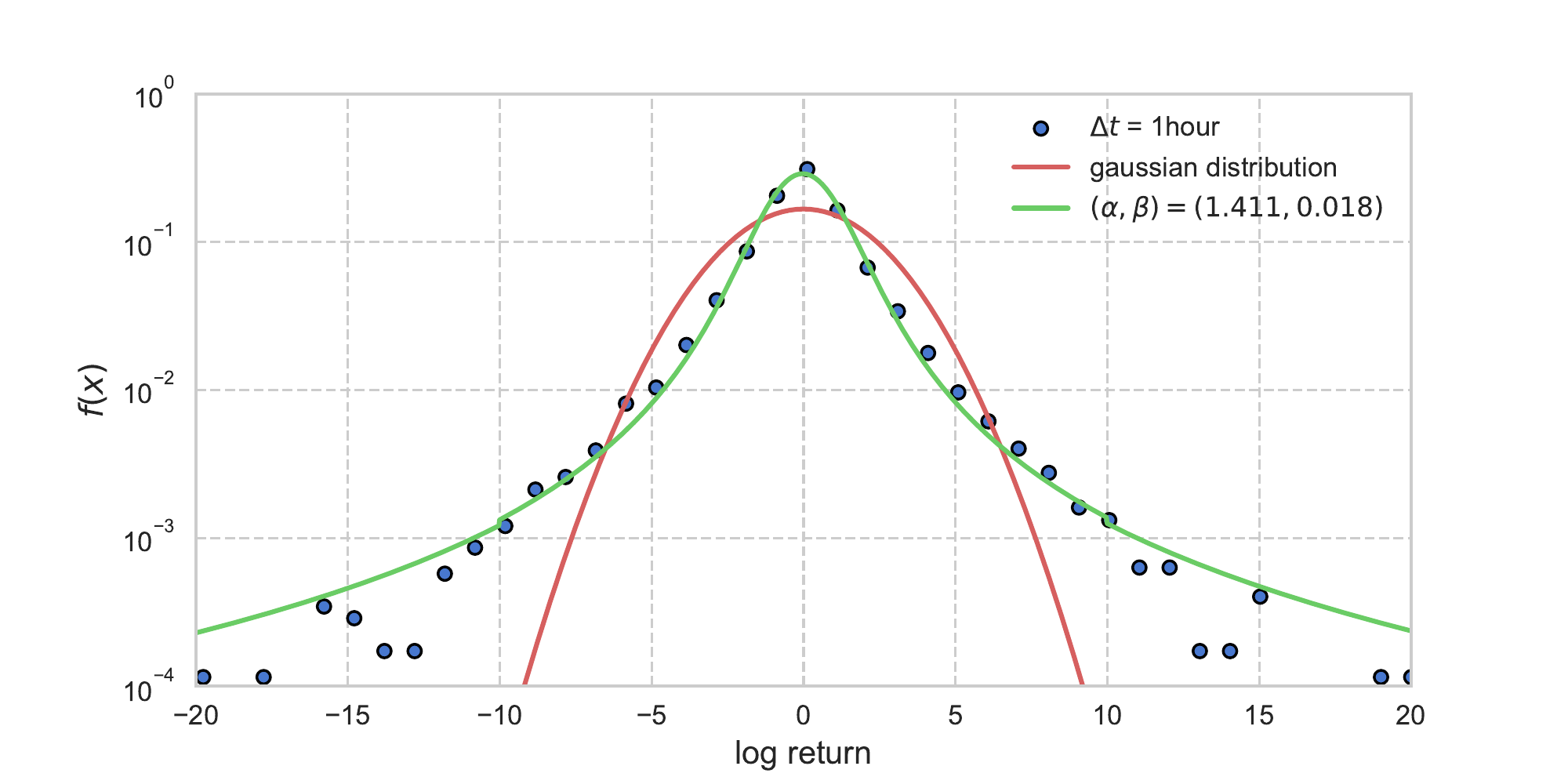}}
%\begin{figure}
 %\centering
  %\subfigure[btc]{\includegraphics[width=\linewidth]{btc_1h_hist.pdf}}
  %\subfigure[ltc]{\includegraphics[width=\linewidth]{ltc_1h_hist.pdf}}
  \caption{ 
   Histogram of standardized empirical data (blue plot) and fitted histogram from estimated stable distribution (green solid line) compared with Gaussian distribution (red solid line).
   Although for significant standardized returns, there is a deficient of data, the estimation well represents the distribution.
   Here, the estimated stable distribution is written by \textit{PyLevy} which is a python package for calculation of PDF for stable distributions.
   (https:\slash\slash{}pylevy.readthedocs.io\slash{}en\slash{}latest\slash{}index.html)
   Note that this computation is supported by the the Maximum Likelihood technique.
   }
   \label{hist}
\end{figure}

We estimate the parameters of the returns for every five currencies over the period from 01/01/2017 to 01/01/2019.
Note that the data set here is every 1-hour data (N=17520).
Cryptocurrency price data are obtained from {\it poloniex} (https://poloniex.com), with all the price exchange rates against USDT.
Here, USDT is an abbreviation of Tether USD, a cryptocurrency asset that maintains the same price and value as the legal US dollar.
For each currency, log-returns (usually called returns) are firstly calculated from the price ${Y_t}$ as $X_t$=$\log Y_{t+\Delta t}-\log Y_t$, where $\Delta t$ is the time interval.
It is then standardized to $(\gamma, \delta)$=$(1,0)$ for easier estimation of parameters $\alpha$ and $\beta$.
Note that the standardization is based on the method in Appendix A.
For $\alpha$, the traditional method discovers that local tails fit an exponent of $\alpha \simeq 2.0 \sim 2.5$, especially for the Bitcoin market \cite{Begusic2018}, however, if we consider fitting a stable distribution, we find different results.
Table~\ref{tb:estimation} shows that the tail index parameter $\alpha$ are estimated roughly in between $1.3$ and $1.5$.
The values are undoubtedly smaller than the $\alpha$=$2$ Gaussian distribution, which indicates that cryptocurrency asset returns are universally {\it non-Gaussian} with fat tails.
Figure~\ref{hist} shows the fitted histogram using the stable distribution for Bitcoin and Litecoin as an example.
The estimated stable distribution well characterizes the fat-tail behaviors and the bulk portion of cryptocurrency asset return distributions, as well as observed in other assets \cite{Mandelbrot1963, MantegnaStanley1995} and financial index \cite{Sandro2018}.
It is worthy of mentioning that Bitcoin and Ripple appear to have $\alpha$ smaller than the other currencies, which is consistent with its fluctuation with prices skyrocketing and falling heavily at the beginning of 2017.
\begin{table}
 \begin{center}
 \caption{Parameter estimation of stable laws for cryptocurrency series with 1-hour time interval data. 
 (2017/01/01-2019/01/01)}
   \begin{tabular}{l p{0mm} c p{15mm} r} \hline
     Cryptocurrency (/USDT) & & $\alpha$ & & \multicolumn{1}{c}{$\beta$} \\ \hline \hline
     Bitcoin (BTC) & & $1.327$ & & $-0.028$ \\
     Ethereum (ETH) & & $1.403$ & & $0.005$ \\
     Ripple (XRP) & & $1.340$ & & $-0.002$ \\
     Litecoin (LTC) & & $1.411$ & & $0.018$ \\
     Monero (XMR) & & $1.518$ & & $0.007$ \\
     \hline
  \end{tabular}
  \label{tb:estimation}
  \end{center}
\end{table}

Both parameters, $\alpha$ and $\beta$, can offer clues to explain the properties of returns.
However, $\beta$ is not so robust to large price fluctuations and tends to have significant estimation errors.
Still, the estimated $\beta$ is close to 0, which means that returns are not so skewed.
The results provide additional views that price fluctuations for cryptocurrency markets exhibit a {\it symmetric} behavior, which is also our finding.
In this paper, we focus on the tail index parameter $\alpha$, which refers to the measure of the tail behaviors and helps further applications of numerical analysis.
%%%%%%%%%%%%%%%%%%%%%%%%%%%%%%%%%%%%%%%%%%%%%%%%%%%%%
% 3.3 Time scaling behavior of Cryptocurrency market
%%%%%%%%%%%%%%%%%%%%%%%%%%%%%%%%%%%%%%%%%%%%%%%%%%%%%
\subsection{Time scaling behavior of Cryptocurrency}
We have argued that in addition to the cryptocurrency market having non-Gaussian features as observed in other financial markets, stable distribution seems to characterize returns reasonably well for a fixed time interval.
This subsection focuses on analyzing cryptocurrency returns with different time scales in order to further understand its behavior.
Meanwhile, we discover the {\it limitations} of the stable distribution for modeling returns in the latter half of this subsection.

Since the analysis of price fluctuations can be done at various time intervals, we go into various time scales.
We use the same datasets for five currencies mentioned in the previous subsection.
The details for time intervals $\Delta t$ are given as follows: 5 minutes, 15 minutes, 30 minutes, 1 hour, 2 hours, 4 hours, 8 hours, and intraday.
With a fixed length of the observation period, the number of data is inversely proportional to $\Delta t$: 210240, 70080, 35040, 17520, 8759, 4379, 2190, and 729, respectively.

We show results in Table \ref{tb:gauss_distance} from the fitting of stable distribution and Gaussian distribution to each of these datasets using the distance measurement.
We first estimate the parameters for stable distributions and use them to obtain the distance, as shown in equation~\eqref{eq:second}.
Since the distance is calculated via the CF expression, each Gaussian distribution is not estimated from the mean and standard deviation but by setting the parameter to $\alpha=2$, and using the estimates $\hat{\gamma}, \hat{\delta}$ ($\beta$ does not necessarily need to be zero, because the CF does not depend on $\beta$ when $\alpha$ is 2).
By doing such numerical assessment, we confirm that stable distribution fit returns {\it better} than the Gaussian distribution--- for {\it all} cases of currencies and time interval conditions.
We also figure out impressive results that both forms of distribution share roughly the same $\Delta t$ with the smallest calculated distance.

\begin{figure}
 \centering
  \includegraphics[width=\linewidth]{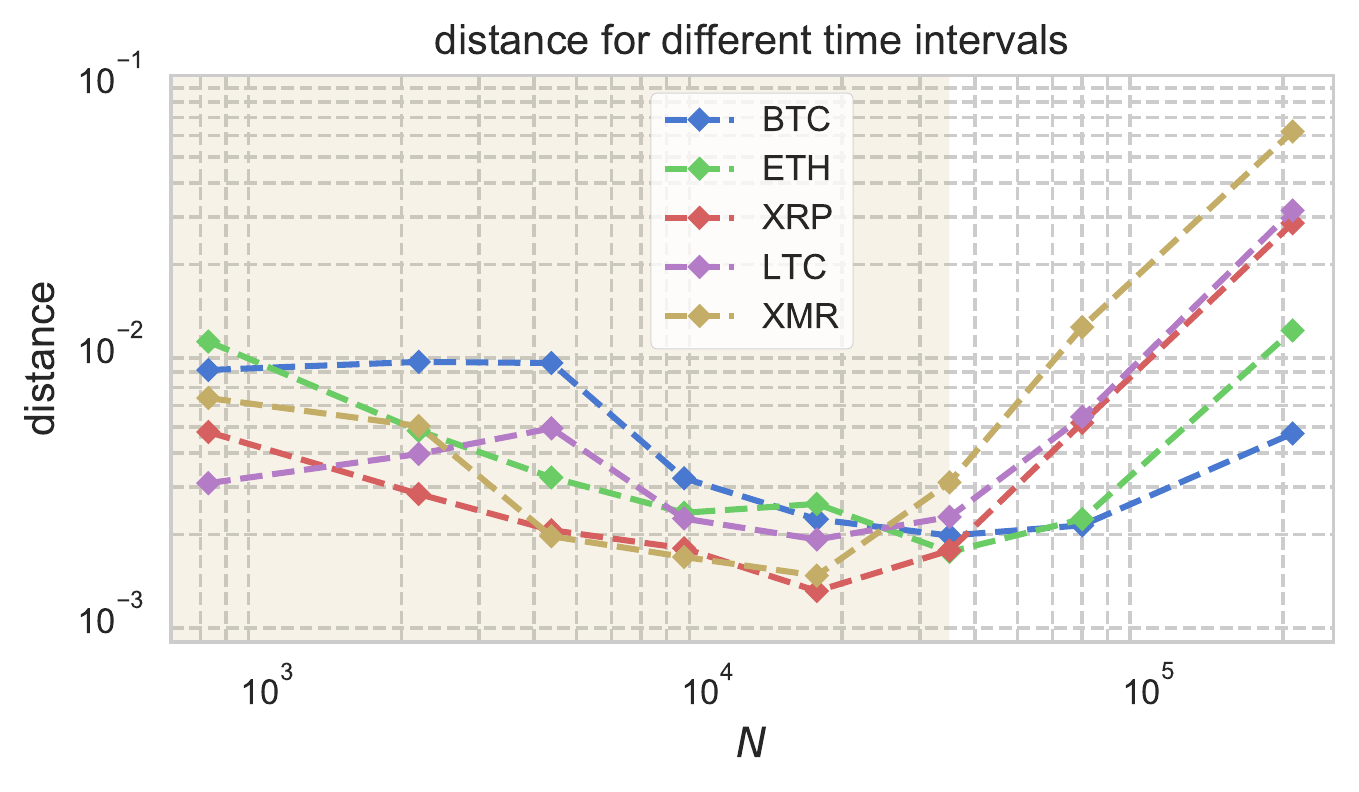}
  \caption{
  The values of distance defined in section 2.3 are shown for different time intervals $\Delta t$.
  The distance decreases with the order $\mathcal{O}\left(1/\sqrt{N}\right)$, but not with the order $\mathcal{O}(1/N)$ due to non-ideal conditions of observed data.
  For a range of intervals shorter than 30 minutes ($\Delta t<30$ min; outside the range in yellow), the distance value increases considerably; this contradicts the idea that the value should decrease (deviation error should also decrease) as the interval becomes shorter with a more significant number of data.
  Therefore, outside the yellow range is not plausible for modeling price fluctuations with stable distribution.
  }
  \label{distance_final}
\end{figure}

\begin{table}
 \begin{center}
 \caption{
 Calculated distance between the empirical distribution and the estimated stable distribution (top row), with calculated distance between the empirical distribution and the estimated Gaussian distribution (bottom row).
 The minimum distance value for each currency is shown in bold for each form of distribution, respectively.
 }  
  \begin{tabular}{l p{0mm} c p{0mm} c p{0mm} c p{0mm} c p{0mm} c} \hline
     $\Delta t$ & & BTC & & ETH & & XRP & & LTC & & XMR\\ \hline \hline
     & & \multicolumn{9}{c}{Estimated stable distribution ($\alpha<2$) ($\times 10^{-3}$)}\\ \cline{3-11}
     5 $\mathrm{min}$ & & $4.73$ & & $11.39$ & & $28.44$ & & $31.68$ & & $62.14$\\
     15 $\mathrm{min}$ & & $2.16$ & & $2.28$ & & $5.18$ & & $5.45$ & & $11.72$\\
     30 $\mathrm{min}$ & & $\textbf{1.98}$ & & $\textbf{1.72}$ & & $1.75$ & & $2.32$ & & $3.12$\\
     1 $\mathrm{hour}$ & & $2.28$ & & $2.59$ & & $\textbf{1.23}$ & & $\textbf{1.92}$ & & $\textbf{1.41}$\\
     2 $\mathrm{hours}$ & & $3.23$ & & $2.41$ & & $1.78$ & & $2.29$ & & $1.65$\\
     4 $\mathrm{hours}$ & & $8.63$ & & $3.25$ & & $2.07$ & & $4.94$ & & $1.97$\\
     8 $\mathrm{hours}$ & & $8.71$ & & $4.85$ & & $2.82$ & & $3.97$ & & $5.04$\\
     1 $\mathrm{Day}$ & & $8.13$ & & $10.36$ & & $4.79$ & & $3.10$ & & $6.40$\\ \hline
      & & \multicolumn{9}{c}{Estimated Gaussian distribution ($\times 10^{-3}$)}\\ \cline{3-11}
     5 $\mathrm{min}$ & & $11.32$ & & $18.32$ & & $44.86$ & & $46.94$ & & $93.22$\\
     15 $\mathrm{min}$ & & $\textbf{8.44}$ & & $7.09$ & & $12.30$ & & $11.60$ & & $17.08$\\
     30 $\mathrm{min}$ & & $8.53$ & & $\textbf{6.65}$ & & $7.86$ & & $7.49$ & & $6.83$\\
     1 $\mathrm{hour}$ & & $9.50$ & & $8.56$ & & $\textbf{7.18}$ & & $7.35$ & & $\textbf{4.80}$\\
     2 $\mathrm{hours}$ & & $10.30$ & & $8.11$ & & $8.26$ & & $8.21$ & & $5.57$\\
     4 $\mathrm{hours}$ & & $16.89$ & & $9.41$ & & $9.65$ & & $11.25$ & & $5.66$\\
     8 $\mathrm{hours}$ & & $14.57$ & & $9.00$ & & $9.87$ & & $9.31$ & & $8.58$\\
     1 $\mathrm{Day}$ & & $13.74$ & & $15.44$ & & $13.10$ & & $\textbf{6.20}$ & & $8.89$\\
     \hline
  \end{tabular}
  \label{tb:gauss_distance}
  \end{center}
\end{table}
We next check the validity of the calculated distance against the stable model.
If observed data entirely agrees with the stable distribution, and if we have unbiased parameter estimates,
the distance value should be close to $1/N$ with deviation error with the order $\mathcal{O}(1/N)$, as discussed in section 2.3.
However, Figure~\ref{distance_final} shows that the calculated distances are likely to be quite above the assumed distance.
The results indicate that observed data is not ultimately consistent with the stable distribution to a complete degree.

One crucial point of issue is that stable distribution presents infinite variance.
This point contradicts the fact that the variance of return for empirical observations turn out to be finite and supports the presence of a finite second moment (local tails appear to be $\alpha \geq 2.0$) \cite{Grabchak2010, Begusic2018}.
Moreover, in the classical study of Mandelbrot \cite{MantegnaStanley1995}, stable distribution appears to fit the empirical return distribution well in the bulk part, but in the very far tails, it seems to overestimate for the sake of its infinite variance.
Strictly speaking, far tails are fatter than those of the empirical return distribution.
This observation is the same for cryptocurrency market--- actual price fluctuations do not show return values {\it too} large (for instance, the largest fluctuation for Bitcoin is 26.9\%), whereas random stable variables include unrealistic extreme values (100\% or 200\% or even larger fluctuations).
This observation may be explained by the causes and effects of the system built in the mechanisms in financial markets, such as the circuit breaker system.
Besides such causes, cryptocurrency prices differ between exchanges.
Those attributes are factors outside the natural fluctuation behaviors but may affect the data we obtain to some extent.
Taking this standpoint gives reasonable assumption to consider that observed data is unfortunately somewhat uncertain, and not ideally perfect to be explained by a stable distribution.
Still, many empirical studies suggest the stable model as a model to examine non-Gaussian behaviors for asset returns because it has solid theoretical reasons to reveal relationships of large and small terms \cite{Xu2011, Bibalan2017, Sandro2018, Chronis2016, Scalas2007}.
As long as stable distribution shows the potential to describe cryptocurrency returns, it is essential to understand the possibilities and limitations of stable models, and to what extent the model is applicable.

If we suppose that unexpected impacts are included in observed data as {\it noise}, one possible stopgap approach to evaluate the distance including bias and noise effects can be given as
\begin{align}
\label{eq:fourth}
    \nonumber
	&\frac{1}{2 \pi} \int_{-\pi}^{\pi}\left|\hat{\phi}(k)-\phi(k)+\varepsilon_{k}\right|^{2} d k \\ \nonumber
	=&\frac{1}{2 \pi} \int_{-\pi}^{\pi}\left|\hat{\phi}(k)-\phi(k)\right|^{2} d k \\ \nonumber
	+&\frac{1}{2 \pi} \int_{-\pi}^{\pi}\left\{2 \operatorname{Re}\left(\left(\hat{\phi}(k)-\phi(k)\right) \overline{\varepsilon_{k}}\right)\right\} d k \\
	+&\frac{1}{2 \pi} \int_{-\pi}^{\pi}\left|\varepsilon_{k}\right|^{2} d k,
\end{align}
where $\hat{\phi}(k)$ is the CF based on estimated parameters, $\phi(k)$ is the true value of the CF, and $\varepsilon_k$ represents the error value of the CF with respect to $k$.
Here, we assume that the empirical CF, $\phi_N(k)$, is equal to the addition of errors and the real CF, $\phi(k)$.
The first term on the right hand of equation~\eqref{eq:fourth} represents distance error caused by parameter estimation errors, the second term is the cross-term between $\hat{\phi}(k)-\phi(k)$ and $\varepsilon_k$, and the last term represents the error caused by noise.
If data is ideal, noise is only related to the random nature of the sampling process.
This case is equivalent to the condition discussed in section 2.3, and it is evident that the distance decreases with the order $\mathcal{O}(1/N)$.
Whereas if noise is large enough to affect the errors $\varepsilon_k$, the cross-term errors (second term) cannot be ignored as well as the errors in the last term.
It is known that
\begin{align*}
	\sqrt{N}\left(\phi_{N}(k)-\phi(k)\right)
\end{align*}
is asymptotically normal with mean 0 \cite{Feuerverger1981}, and hence the cross-term decreases with the order $\mathcal{O}\left(1/\sqrt{N}\right)$.
The leading order of the whole distance equation then becomes $\mathcal{O}\left(1/\sqrt{N}\right)$ for large $N$.
Therefore, if we admit some observational noises, it may be reasonable and acceptable that the distances of empirical studies are larger than the natural distance error $1/N$.
If we have biased parameter estimates for $\hat{\phi}(k)$, we will have even more considerable distances.
However, it should be noticed in Figure~\ref{eq:third} that even though we consider unexpected noise in cryptocurrency data, we find numerical evidence of stable distributions fail to capture returns under some conditions of time intervals.
The results show that for high-frequency data ($\Delta t$ shorter than 30 minutes), the distance values become too large.
In other words, when we have more abundant data available, distances tend to increase, which implies that stable distribution models do not perform well with $\Delta t$ shorter than 30 minutes.
Some of the possible causes of the distance separation could be, for instance, the scarcity of tick-to-tick fluctuation patterns on less active exchanges and the market microstructure noise seen in high-frequency data \cite{Ait2010}.
What is more, data may no longer be stationary with overly high-frequent conditions owing to the volatility clustering phenomenon.
If data is non-stationary, {\it ergodicity} does not hold.

\begin{figure}
 \centering
  \includegraphics[width=\linewidth]{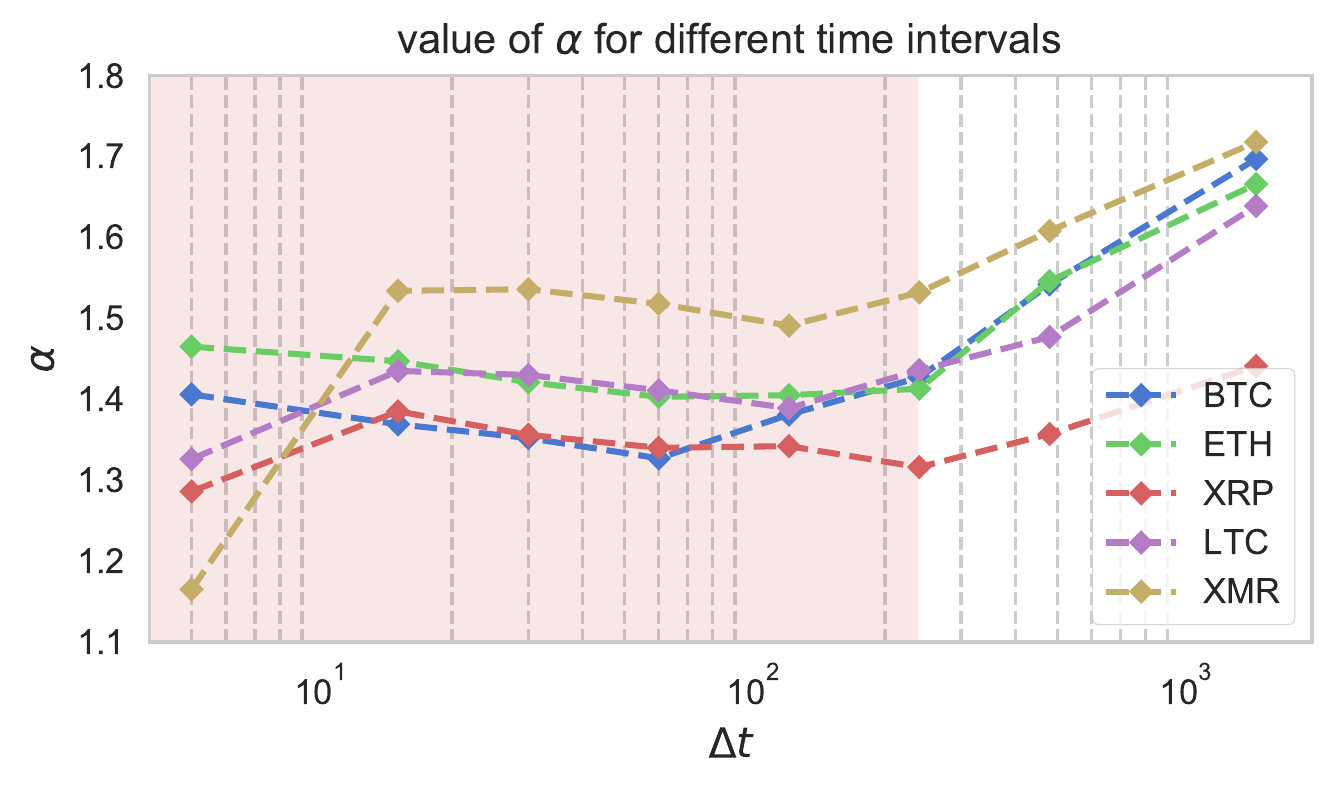}
  \caption{
  The estimate value of $\alpha$ are shown for different time intervals $\Delta t$ in order to investigate the property of time scaling.
  While the stable regime holds for intervals smaller than 4 hours ($\Delta t< 4$ hours; range in pink), for longer time scales, $\alpha$ tends to increase towards $\alpha=2$: the Gaussian regime.
  Although the crossover between the two regimes seem to take slightly different values depending on the choice of cryptocurrency, $\Delta t = 4$ hours is a good agreement.
  \label{distance2}
}
\end{figure}

In Mandelbrot's pioneering investment of cotton prices \cite{Mandelbrot1963}, he observed that in addition to being non-Gaussian, returns show another endogenous interest--- the invariant property of \textit{time scaling}, which means that the return distribution for every various time interval $\Delta t$ potentially show a similar class of functions conforming to a stable distribution.
This behavior is certainly well connected with the Generalized Central Limit Theorem (GCLT), and hence the idea of exploring the scaling behavior is natural and essential when modeling financial assets with stable distributions.
Mandelbrot has discovered that $\Delta t$ ranging from 1 day up to 1 month shows consistent forms with the stable distribution.
Gopikrishnan et al. \cite{Gop1999} also studied another asset of the S\&P 500 index, showing that the distribution for $\Delta t$ smaller than $4$ days have consistent forms as well.
In a trivial sense, however, not all ranges of $\Delta t$ show excellent compatibility with stable distributions.
In addition, these previous studies show that financial asset returns tend to have less fat tails when analyzed with long timescales.
This is because finite empirical observations do not support the GCLT, and the scaling property does not hold for long timescales but converges to a Gaussian distribution by the Central Limit Theorem (CLT).

To overcome these issues, Mantegna et al. were the first to propose the Truncated L\'evy Distribution (TLD) \cite{Mantegnatruncatecut1994}.
The central part of the TLD is consistent with the stable distribution, but its far tail has a discrete cutoff.
Koponen improved the TLD by introducing a smooth exponential cutoff to make it possible to derive an analytic expression for the CF and easier computation simulations \cite{Koponen1995}.
For both cases, far tails have a faster decay compared to the stable laws. This assures the variance to be finite, and fortunately, more or less preserves the stable properties.
This development can be explained when we consider the sums of independent and identically random variables following the TLD, known as the Truncated L\'evy Flight (TLF).
Since most of the distribution is like a stable distribution but has a finite variance, the TLF process \textit{converges slowly} to the Gaussian distribution \cite{Mantegnatruncatecut1994, Koponen1995}.
For relatively short timescales, the influence of the truncated tails is too slight to affect the stochastic process for the CLT to be applied.
It does not converge to the expected Gaussian distribution, but still under the GCLT.
Therefore, as long as the stable regime holds, such a stochastic process can approximately be expressed as a stable distribution.
Once the process reaches the crossover, it starts to go towards the Gaussian behavior.
Overall, stable processes accounting for TLD provides two forms of distributions in terms of time scaling: {\it stable regime} and {\it Gaussian regime}.

Figure~\ref{distance2} displays the shift of $\alpha$ for different time scaling in cryptocurrency markets, which enables us to discuss the correspondence between the empirical behavior and the theoretical background of the GCLT.
The results imply that the stable model is inappropriate for $\Delta t$ larger than 4 hours, where the GCLT is not valid, and the stable regime does not hold.
We have reported in Figure~\ref{distance_final} that in the case of high-frequency data ($N>35040; \Delta t < 30\mathrm{min}$), distance results are undoubtfully too large to support the model.
If we consider the presence of observational noises in data, a range of approximately $30 \mathrm{min}\leq \Delta t \leq 4 \mathrm{h}$ seems to be moderate for analyzing cryptocurrency returns when employing stable distributions.
In this range, the return distributions satisfy the stable process with
relatively small distance values.
%%%%%%%%%%%%%%%%%%%%%%%%%%%%%%%%%%%%%%%%%%%%%%%%%%%%%
% 3.4 Performance of stable law fit and alternative one-sided distribution models
%%%%%%%%%%%%%%%%%%%%%%%%%%%%%%%%%%%%%%%%%%%%%%%%%%%%%
\subsection{Performance of stable law fit and alternative distribution models}
The approaches explained in the previous subsection contribute to demonstrating how to apply the stable model properly.
However, the approaches do not necessarily identify the actual model that describes the fluctuation system.
An alternative model may be more appropriate even when the conditions, including time scaling, for supporting the stable model are satisfied.
As we have mentioned before, the infinite properties of stable distributions make it challenging to build a `good' modeling for the far tail portions.
The power-law model is widely accepted, mainly when focusing on the tails, and often rules out the stable regime in empirical data with finite variance.

To identify the appropriate model for fitting cryptocurrency data, we employ the {\it likelihood ratio test} explained in subsection 2.4.
We compare the stable model with each of the two alternative models of one-sided distributions: power-law and exponential distributions.
We first focus on the identical range of local tails for the two distributions, since the comparison should be made under the identical tail conditions.
As explained in subsection 2.2, the tail is defined as the one that shows the best fit with the alternative distribution in terms of KS statistics.
Tables~\ref{tb:powerexp_basic_right} and~\ref{tb:powerexp_basic_left} of Appendix D show the results of fitting the data of standardized returns with each of the two alternative distributions for the positive and the negative tails, respectively.
Appendix D provides some discussions for the model fitting.
Tables~\ref{tb:powerexp_test_right} and~\ref{tb:powerexp_test_left} show the results of the likelihood ratio tests (the likelihood ratio $\mathcal{R}$ and the corresponding $p$-value) under several tail portions for the positive and the negative tails of the standardized empirical returns, respectively.
Time intervals $\Delta t=1$ hour and $\Delta t=2$ hours are examined, where the time scaling conditions are well satisfied, as presented in subsection 3.3.
It should be noted that the standard density and distribution functions of the stable distribution are numerically derived approximately by implementing the Fourier integral formulas~\cite{Zolotarev1986, Nolan1997}, which are available in package {\it libstable}~\cite{Roy2017}.
We set aside any issues related to numerical approximations of the stable distributions since we aim to directly compare the fitting between models rather than the assessments of a specific model.
Under the estimated tails (columns $\hat{x}_\mathrm{min}$ in the Tables~\ref{tb:powerexp_test_right} and~\ref{tb:powerexp_test_left}), the values of $\mathcal{R}$ are negative for most cases, meaning that alternative distributions achieve a better fit than the stable distribution.
Moreover, the results in Appendix D imply that empirical returns in the estimated tails are plausible with power laws.
These are consistent with the arguments in many empirical studies that local tails of financial returns often exhibit the inverse cubic law.

However, our primary objective is to evaluate the entire or a wider range of returns.
Since the estimated tail portions sometimes leave too many observations out of consideration, analyzing the data covering more observations, including the estimated tail portions, is needed to capture the characteristics of the data in a more comprehensive manner.
In addition, we can explain the advantages of stable distribution only when the entire distribution is considered.
We select the largest and lowest 5\%, 15\%, 30\%, and 45\% portions of the data as the lower and upper bound of tails, respectively, in order to reveal which distribution shows a better fit for larger portions of tails.
The likelihood ratios for the 5\% tail portion show negative values for most cases, like those with the estimated tails.
However, as the tail portions become large, the likelihood ratios generally turn to positive values, particularly for the tail portions larger than 15\%.
Although some results are statistically insignificant, we confirm that the stable model tends to present a better fitting of returns for large portions of tails.
Our results indicate that when we focus on the tails, for example, investigating the tail risks, the power-law or exponential model is suggested.
But when we examine the characteristics of cryptocurrency price fluctuations from the entire data, the stable distribution is suggested to be an appropriate model for the analysis of behavior issues.
A perfect model to explain the behaviors remains to be a challenging issue, but our strategies of coping with stable models are helpful for any extension of the model.
\begin{table*}
\centering
\caption{
The {\it likelihood ratio tests} for the {\it positive} tail of standardized cryptocurrency returns.
The tests compare the stable model with each of the two alternative models of one-sided distributions: power-law and exponential distributions, for the tail behavior.
In addition to the estimated tails with lower bound $\hat{x}_\mathrm{min}$, we select the largest 5\%, 15\%, 30\%, and 45\% portions of the data as the lower bound of tails, respectively.
For each tail portion, the stable distribution is compared with the best fit with alternative distributions in terms of KS statistics.
Note that the empirical fit with the exponential model could be estimated using the same idea as the power-law model, in which the method relies on the {\it maximum likelihood estimator} and the KS statistics (subsection 2.2, Appendix D).
The likelihood ratios $\mathcal{R}$ in equation~\eqref{eq:likelihood} and the corresponding $p$-values are given for each case.
If the values are smaller than 0.1, we say that the calculated ratios $\mathcal{R}$ are statistically significant (shown in bold).
}
\begin{tabular}{l l p{0mm} c c p{0mm} c c p{0mm} c c p{0mm} c c p{0mm} c c}
\hline
\multicolumn{2}{l}{Power law} & & \multicolumn{2}{c}{$\hat{x}_\mathrm{min}$} & & \multicolumn{2}{c}{5\%} & & \multicolumn{2}{c}{15\%} & & \multicolumn{2}{c}{30\%} & & \multicolumn{2}{c}{45\%} \\
\cline{4-5} \cline{7-8} \cline{10-11} \cline{13-14} \cline{16-17}
dataset & $\Delta t$ & & $\mathcal{R}$ & p & & $\mathcal{R}$ & p & & $\mathcal{R}$ & p & & $\mathcal{R}$ & p & & $\mathcal{R}$ & p \\ \hline \hline
BTC & 1h & & -87.54 & \textbf{0.000} & & -93.95 & \textbf{0.000} & & -6.736 & 0.209 & & 432.7 & \textbf{0.000} & & 2201 & \textbf{0.000} \\
 & 2h & & -45.84 & \textbf{0.000} & & -38.33 & \textbf{0.000} & & 32.02 & 0.196 & & 185.7 & \textbf{0.000} & & 980.3 & \textbf{0.000} \\
ETH & 1h & & -42.13 & \textbf{0.000} & & -77.84 & \textbf{0.000} & & 24.56 & \textbf{0.000} & & 445.8 & \textbf{0.000} & & 2864 & \textbf{0.000} \\
 & 2h & & -37.31 & \textbf{0.000} & & -36.33 & \textbf{0.000} & & 2.778 & 0.411 & & 219.7 & \textbf{0.000} & & 1377 & \textbf{0.000} \\
XRP & 1h & & -42.58 & \textbf{0.000} & & -42.69 & \textbf{0.000} & & 2.361 & 0.495 & & 330.4 & \textbf{0.000} & & 2683 & \textbf{0.000} \\
 & 2h & & -25.70 & \textbf{0.000} & & -21.32 & \textbf{0.000} & & 0.638 & 0.788 & & 157.5 & \textbf{0.000} & & 1032 & \textbf{0.000} \\
LTC & 1h & & -50.00 & \textbf{0.000} & & -63.00 & \textbf{0.000} & & 25.94 & \textbf{0.000} & & 410.1 & \textbf{0.000} & & 2940 & \textbf{0.000} \\
 & 2h & & -31.15 & \textbf{0.000} & & -33.21 & \textbf{0.000} & & 2.450 & 0.434 & & 228.7 & \textbf{0.000} & & 1362 & \textbf{0.000} \\
XMR & 1h & & -16.36 & \textbf{0.000} & & -41.29 & \textbf{0.000} & & 18.78 & \textbf{0.000} & & 491.3 & \textbf{0.000} & & 3370 & \textbf{0.000} \\
 & 2h & & -27.12 & \textbf{0.000} & & -22.78 & \textbf{0.000} & & 9.027 & \textbf{0.009} & & 229.1 & \textbf{0.000} & & 1321 & \textbf{0.000} \\ \hline
\multicolumn{2}{l}{Exponential} & & \multicolumn{2}{c}{$\hat{x}_\mathrm{min}$} & & \multicolumn{2}{c}{5\%} & & \multicolumn{2}{c}{15\%} & & \multicolumn{2}{c}{30\%} & & \multicolumn{2}{c}{45\%} \\
\cline{4-5} \cline{7-8} \cline{10-11} \cline{13-14} \cline{16-17}
dataset & $\Delta t$ & & $\mathcal{R}$ & p & & $\mathcal{R}$ & p & & $\mathcal{R}$ & p & & $\mathcal{R}$ & p & & $\mathcal{R}$ & p \\ \hline \hline
BTC & 1h & & -108.3 & \textbf{0.000} & & -109.6 & \textbf{0.000} & & -79.02 & \textbf{0.000} & & -24.67 & 0.393 & & 12.03 & 0.700 \\
& 2h & & -41.90 & \textbf{0.000} & & -42.95 & \textbf{0.000} & & -38.52 & \textbf{0.013} & & -13.65 & 0.503 & & -4.713 & 0.826 \\
ETH & 1h & & -46.52 & \textbf{0.000} & & -67.93 & \textbf{0.000} & & -54.41 & \textbf{0.015} & & 2.518 & 0.932 & & 18.37 & 0.560 \\
& 2h & & 11.78 & 0.603 & & -31.7 & \textbf{0.001} & & -9.454 & 0.557 & & 10.20 & 0.627 & & 18.80 & 0.398 \\
XRP & 1h & & -46.37 & \textbf{0.000} & & -37.09 & \textbf{0.002} & & 85.24 & \textbf{0.002} & & 254.8 & \textbf{0.000} & & 323.0 & \textbf{0.000} \\
& 2h & & -17.07 & \textbf{0.095} & & -10.26 & 0.440 & & 48.18 & \textbf{0.063} & & 127.3 & \textbf{0.000} & & 176.6 & \textbf{0.000} \\
LTC & 1h & & -54.55 & \textbf{0.000} & & -51.78 & \textbf{0.001} & & -25.10 & 0.347 & & 63.02 & \textbf{0.074} & & 85.64 & \textbf{0.022} \\
& 2h & & -32.64 & \textbf{0.000} & & -36.6 & \textbf{0.000} & & -7.514 & 0.620 & & 17.26 & 0.404 & & 43.10 & \textbf{0.053} \\
XMR & 1h & & 42.26 & 0.125 & & -41.64 & \textbf{0.001} & & 1.391 & 0.951 & & 25.19 & 0.351 & & 36.49 & 0.184 \\
& 2h & & 6.303 & 0.746 & & -25.90 & \textbf{0.002} & & -6.662 & 0.661 & & 12.27 & 0.509 & & 7.700 & 0.690 \\ \hline
\end{tabular}
\label{tb:powerexp_test_right}
\end{table*}
\begin{table*}
\centering
\caption{
The {\it likelihood ratio tests} for the {\it negative} tail of standardized cryptocurrency returns.
The tests compare the stable model with each of the two alternative models of one-sided distributions: power-law and exponential distributions, for the tail behavior.
In addition to the estimated tails with upper bound $\hat{x}_\mathrm{min}$, we select the lowest 5\%, 15\%, 30\%, and 45\% portions of the data as the upper bound of tails, respectively.
The likelihood ratios $\mathcal{R}$ and the corresponding $p$-values are given for each case.
Statistically significant test are shown in bold.
}
\begin{tabular}{l l p{0mm} c c p{0mm} c c p{0mm} c c p{0mm} c c p{0mm} c c}
\hline
\multicolumn{2}{l}{Power law} & & \multicolumn{2}{c}{$\hat{x}_\mathrm{min}$} & & \multicolumn{2}{c}{5\%} & & \multicolumn{2}{c}{15\%} & & \multicolumn{2}{c}{30\%} & & \multicolumn{2}{c}{45\%} \\
\cline{4-5} \cline{7-8} \cline{10-11} \cline{13-14} \cline{16-17}
dataset & $\Delta t$ & & $\mathcal{R}$ & p & & $\mathcal{R}$ & p & & $\mathcal{R}$ & p & & $\mathcal{R}$ & p & & $\mathcal{R}$ & p \\ \hline \hline
BTC & 1h & & -66.46 & \textbf{0.000} & & -82.95 & \textbf{0.000} & & 15.07 & \textbf{0.001} & & 393.4 & \textbf{0.000} & & 2666 & \textbf{0.000} \\
& 2h & & -48.64 & \textbf{0.000} & & -47.58 & \textbf{0.000} & & 17.19 & \textbf{0.000} & & 221.9 & \textbf{0.000} & & 1830 & \textbf{0.000} \\
ETH & 1h & & -68.01 & \textbf{0.000} & & -68.96 & \textbf{0.000} & & 11.04 & \textbf{0.006} & & 373.8 & \textbf{0.000} & & 2355 & \textbf{0.000} \\
& 2h & & -35.95 & \textbf{0.000} & & -35.04 & \textbf{0.000} & & 10.17 & \textbf{0.000} & & 194.1 & \textbf{0.000} & & 1265 & \textbf{0.000} \\
XRP & 1h & & -54.58 & \textbf{0.000} & & -66.16 & \textbf{0.000} & & -0.432 & 0.936 & &373.7 & \textbf{0.000} & & 2894 & \textbf{0.000} \\
& 2h & & -33.77 & \textbf{0.000} & & -35.37 & \textbf{0.000} & & 1.754 & 0.643 & & 200.3 & \textbf{0.000} & & 1531 & \textbf{0.000} \\
LTC & 1h & & -45.76 & \textbf{0.000} & & -65.09 & \textbf{0.000} & & 14.92 & \textbf{0.000} & & 383.4 & \textbf{0.000} & & 2481 & \textbf{0.000} \\
& 2h & & -35.97 & \textbf{0.000} & & -35.50 & \textbf{0.000} & & 2.819 & 0.402 & & 194.9 & \textbf{0.000} & & 1196 & \textbf{0.000} \\
XMR & 1h & & -53.03 & \textbf{0.000} & & -45.08 & \textbf{0.000} & & 17.79 & \textbf{0.001} & & 406.4 & \textbf{0.000} & & 3068 & \textbf{0.000} \\
& 2h & & -21.26 & \textbf{0.000} & & -24.98 & \textbf{0.000} & & 9.449 & \textbf{0.006} & & 207.8 & \textbf{0.000} & & 1479 & \textbf{0.000} \\ \hline
\multicolumn{2}{l}{Exponential} & & \multicolumn{2}{c}{$\hat{x}_\mathrm{min}$} & & \multicolumn{2}{c}{5\%} & & \multicolumn{2}{c}{15\%} & & \multicolumn{2}{c}{30\%} & & \multicolumn{2}{c}{45\%} \\
\cline{4-5} \cline{7-8} \cline{10-11} \cline{13-14} \cline{16-17}
dataset & $\Delta t$ & & $\mathcal{R}$ & p & & $\mathcal{R}$ & p & & $\mathcal{R}$ & p & & $\mathcal{R}$ & p & & $\mathcal{R}$ & p \\ \hline \hline
BTC & 1h & & -93.17 & \textbf{0.000} & & -94.11 & \textbf{0.000} & & -57.95 & \textbf{0.017} & & 62.24 & \textbf{0.053} & & 147.2 & \textbf{0.000} \\
& 2h & & -42.55 & \textbf{0.000} & & -43.99 & \textbf{0.000} & & -37.16 & \textbf{0.014} & & 16.41 & 0.437 & & 41.28 & \textbf{0.073} \\
ETH & 1h & & -68.45 & \textbf{0.000} & & -74.35 & \textbf{0.000} & & -39.82 & \textbf{0.078} & & 48.10 & 0.109 & & 73.89 & \textbf{0.022} \\
& 2h & & -39.24 & \textbf{0.000} & & -40.38 & \textbf{0.000} & & -31.23 & \textbf{0.034} & & 12.49 & 0.526 & & 28.97 & 0.167 \\
XRP & 1h & & -16.92 & \textbf{0.000} & & -40.72 & \textbf{0.016} & & 39.91 & 0.195 & & 151.2 & \textbf{0.000} & & 190.4 & \textbf{0.000} \\
& 2h & & -17.30 & \textbf{0.000} & & -29.20 & \textbf{0.007} & & 1.102 & 0.954 & & 47.34 & \textbf{0.070} & & 67.88 & \textbf{0.018} \\
LTC & 1h & & -74.24 & \textbf{0.000} & & -73.22 & \textbf{0.000} & & -46.59 & \textbf{0.037} & & 28.25 & 0.334 & & 43.12 & 0.162 \\
& 2h & & -46.98 & \textbf{0.000} & & -47.18 & \textbf{0.000} & & -30.43 & \textbf{0.021} & & -1.53 & 0.931 & & -7.746 & 0.679 \\
XMR & 1h & & -14.74 & 0.635 & & -39.60 & \textbf{0.007} & & -5.292 & 0.828 & & 40.46 & 0.167 & & 44.55 & 0.132 \\
& 2h & & 4.292 & 0.810 & & -29.35 & \textbf{0.000} & & -14.04 & 0.311 & & 9.749 & 0.570 & & 9.988 & 0.570 \\ \hline
\end{tabular}
\label{tb:powerexp_test_left}
\end{table*}
%%%%%%%%%%%%%%%%%%%conclusion
\section{Conclusion}
This paper has explored the behaviors of price fluctuations in cryptocurrency markets by applying the L\'evy's stable distribution 
and discussed its validity for the empirical analysis.
We provide numerical, theoretical, and justifications for supporting the stable distribution as a practical model to understand the fluctuation phenomena in financial systems.

We focus on characterizing the entire dataset of returns, including the tail behaviors.
The stable distribution takes into account the entire observations.
With the use of the proper estimation method, we find that returns exhibit stable laws with tail index $\alpha \simeq 1.4$ and $\beta \simeq 0$ (symmetric).
We introduce a numerical approach based on the CF and a theoretical approach based on the GCLT to find evidence for stable laws, by focusing on the time scaling behavior with different time intervals.
Similar to other financial asset returns, our results of the numerical approach suggest that cryptocurrency price returns follow a fat-tailed stable distribution better than the Gaussian distribution for {\it all} time scales.
However, we find that even if we admit some observational noise terms, the numerical distance shows implausible results for high-frequency data.
On the other hand, the theoretical approach based on the GCLT represents implausible results for low-frequency data, where the stable regime breaks down to a Gaussian regime.
From both points of view, our assessment implies that the stable model is not necessarily acceptable for any analysis condition.
We propose that the combination of these approaches helps understand the intriguing properties of asset fluctuations, and gives us insight into appropriate ranges of time scaling for modeling with stable distribution in a more careful sense.
In particular, a time scaling condition of ranging roughly 30 minutes to 4 hours is concluded to be a suitable range of intervals for cryptocurrency markets, where both quantitative and theoretic properties are consistent with the stable model.

Moreover, we confirm the potency of modeling returns with stable distributions under some time scaling conditions by clarifying which distribution shows a better fit among controversial fat-tailed distributions.
We find statistical evidence that when more than 15\% tail portion of data is considered, the stable distribution dominates other alternative distributions.
The results imply that the stable model is comparatively appropriate for characterizing the entire or broader range of the data.
At the same time, we find that the far tails generally follow a power law, which coincides with the results in many empirical studies on tail behaviors of returns.
Therefore, these ideas can be developed to create some benchmarks for risk management and portfolio theories.
To reach a more rigorous conclusion on whether stable models may work in practical applications, however, a more elaborate discussion would be necessary.
%
%\begin{acknowledgment}
%\acknowledgment
\newline

The authors would like to thank Dr. Shin-itiro GOTO, Kyoto University, for giving us fruitful discussions.

%\end{acknowledgment}

\appendix
\section{Parameter Estimation Methods of Stable Laws}%%%%%%%%%%%%%%%%%%%%%%
The method proposed by Koutrovelis \cite{Koutrouvelis1980} is a straightforward approach to estimate parameters with high accuracy; however, it lacks the procedure of standardization.
The method also needs different values of \textit{Optimum Number of Points} for different estimated $\alpha$, which is unsuitable for applications of $\alpha$.
The method is slightly modified more practically, shown as below \cite{Fukunaga2018}.

When analyzing data, it is common to assume the data are ergodic \cite{ArnoldAvez}.
If $X_n\:(n=1,2,\ldots)$ are ergodic for the measure $\rho(x)\,dx$ in space $M$, the following equation holds \cite{Umeno2016}:
\begin{align*}
	\lim_{N \to \infty} \frac{1}{N} \sum_{n=1}^{N} \mathrm{e}^{ikX_n} &= \int_{M} \mathrm{e}^{ikx} \rho(x)\,dx.
\end{align*}
This ergodic equality indicates that the CF $\phi(k)$ is obtained as
\begin{align*}
	\phi(k) = \lim_{N \to \infty} \frac{1}{N} \sum_{n=1}^{N} \mathrm{e}^{ikX_n}.
\end{align*}
We can calculate the empirical CF $\phi_N(k)$ for a large number of data set $X_n\:(n=1,2,\ldots,N)$ as 
\begin{align}
	\label{eq:1}
	\phi_N(k) = \frac{1}{N} \sum_{n=1}^{N} \mathrm{e}^{ikX_n}.
\end{align}
Since the empirical CF, $\phi_N(k)$, of the L\'evy's stable distribution for ${\alpha \neq 1}$ and ${k > 0}$ can be written as
\begin{align}
	\label{eq:2}
	\phi_N(k) =\exp \left \{ i \delta k -(\gamma k)^{\alpha} \left [1-i \beta \tan \left ( \frac{\pi \alpha}{2} \right ) \right ] \right \},
\end{align}
we obtain the following equations for the empirical CF,
\begin{align}
	\label{eq:3}
	\log(-\log|\phi_N(k)|) &=\alpha \log k + \alpha \log \gamma, \\
	\label{eq:4}
	\frac{1}{k} \arctan \left ( \frac{\phi_{N, I} (k)}{\phi_{N, R} (k)} \right )  &= \beta \gamma^\alpha \tan \left ( \frac{\pi \alpha}{2} \right ) k^{\alpha-1} + \delta.
\end{align}
Here, 
${\phi_{N, R} (k)}$ is the real part, and ${\phi_{N, I} (k)}$ is the imaginary part of the empirical CF.
With ${y:=\log(-\log|\phi_N(k) |)}$ and ${x:=\log k}$, the linear regression of equation
$\eqref{eq:3}$ gives the estimators $\hat{\alpha}$ and $\hat{\gamma}$.
Then, with the obtained estimators $\hat{\alpha}$, $\hat{\gamma}$, ${y=(1/k) \arctan \{ \phi_{N, I} (k) / \phi_{N, R} (k) \}}$ and ${x=\hat{\gamma}^{\hat{\alpha}} \tan(\pi \hat{\alpha} / 2) k^{\hat{\alpha}-1} }$, the linear regression of equation
$\eqref{eq:4}$ gives the estimators ${\hat{\beta}}$ and ${\hat{\delta}}$.
Note that $k\rightarrow0$ for CF corresponds to $x\rightarrow \infty$ for the PDF.
Since the characteristics of $\alpha$ appear in the tail portion of the PDF, the range for regression is $k=[0.2,1.0]$ with bins of 0.01.

Standardization is an important process for estimation since it is tough to estimate the parameters accurately when the scale and the location parameters $(\gamma, \delta)$ are far from the standardized $(1,0)$.
The procedure is supported by the property of L\'evy's stable distribution shown as below:

When the random variables are $X \overset{\mathrm{d}}{=} S(\alpha, \beta, 1, 0)$, the transformed variables $Y$ with $\gamma' > 0$ and $\delta' \in (-\infty, \infty)$,
\begin{align*}
	Y := \gamma' X + \delta'
\end{align*}
also satisfy $X \overset{\mathrm{d}}{=} S(\alpha, \beta, \gamma', \delta')$.
According to this property, the process for standardization can be obtained.
After the estimation of $\hat{\gamma}$ from equation $\eqref{eq:3}$, rescale $X$ by
\begin{align*}
	X:= \frac{X}{\hat{\gamma}},
\end{align*}
until ${1-\epsilon < \hat{\gamma} < 1+\epsilon}$ is satisfied.
Next, estimate $\hat{\delta}$ from equation $\eqref{eq:4}$ and relocate $X$ by
\begin{align*}
	X:= X-\hat{\delta},
\end{align*}
until ${-\epsilon < \hat{\delta} < \epsilon}$ is satisfied.

When it is obvious that the results show $\alpha > 1$, the mean of the L\'evy's stable distribution turns equal to the location parameter.
So, in this case, it is also possible to estimate $\hat{\delta}$ more simply by
\begin{align*}
	\hat{\delta}=\frac{1}{n} \sum_{i=1}^n X_i. \\
\end{align*}
%Stable Simulations
\section{Stable Simulations}%%%%%%%%%%%%%%%%%%%
There are several algorithms to generate the sequence of L\'evy's stable distribution, such as the classical method by Chambers \cite{Chambers1976} and the method based on the superposition of chaotic processes by Umeno \cite{Umeno1998}.
Weron \cite{Weron1996} has made a few corrections to the Chamber's algorithm with the fastest in calculation, which provides a simple algorithm shown as the following:

Generate a random variable $V$ uniformly distributed on $(-\frac{\pi}{2}, \frac{\pi}{2})$ and an independent exponential random variable $W$ with mean 1.
\begin{enumerate}
\item if $\alpha \neq 1$, calculate \\
\begin{align*}
	X &= S_{\alpha, \beta} \times \frac{\sin{(\alpha(V+B_{\alpha, \beta}))}}{(\cos(V))^{\frac{1}{\alpha}}} \\
	& \times \left ( \frac{ \cos{(V-\alpha(V+B_{\alpha, \beta}))}}{W} \right ) ^{\frac{1-\alpha}{\alpha}},
\end{align*}
where
\begin{align*}
	B_{\alpha, \beta} &= \frac{\arctan(\beta \tan{\frac{\pi \alpha}{2}})}{\alpha}, \\
	S_{\alpha, \beta} &= \left ( 1+\beta^2 \tan^2 \frac{\pi \alpha}{2} \right )^{\frac{1}{2\alpha}}.
\end{align*}
\item if $\alpha = 1$, calculate \\
\begin{align*}
	X=\frac{2}{\pi} \left [ \left ( \frac{\pi}{2}+\beta V \right ) \tan V - \beta \log \left ( \frac{\frac{\pi}{2}W \cos V}{\frac{\pi}{2}+\beta V} \right ) \right ].
\end{align*}
\end{enumerate}
By generating $W$ and $V$ a sufficient number of times, the formula allows to construct a standardized L\'evy stable random variable ${X \overset{\mathrm{d}}{=}S(\alpha, \beta, 1,0)}$ for $\alpha \in (0,2]$ and $ \beta \in [-1,1]$.
%Calculation for the expectation of the distance function
\newpage
\section{Calculation for the Expectation of the Distance Function}%%%%%%%%%%%%
When we consider data $X_n (n=1,2,\cdots,N)$ to be i.i.d. distributed from some certain distribution, and $\phi_N(k)$ becomes the true value $\phi(k)$ as $N\rightarrow \infty$, we obtain
\begin{align*}
	E\left[ \left | \phi_N (k) \right |^2 \right ] 
	&= E\left[ \left( \frac{1}{N} \sum_{n=1}^N \mathrm{e}^{ikX_n} \right) \overline{\left( \frac{1}{N} \sum_{n=1}^N \mathrm{e}^{ikX_n} \right)} \right] \\
	&= \frac{1}{N^2} E\left[ N+ \sum_{\substack{n,l=1\\n\neq l}}^N \mathrm{e}^{ikX_n} \overline{\mathrm{e}^{ikX_l}} \right] \\
	&= \frac{1}{N^2} \left( N+ \sum_{\substack{n,l=1\\n\neq l}}^N E \left[ \mathrm{e}^{ikX_n} \overline{\mathrm{e}^{ikX_l}} \right ] \right) \\
	&= \frac{1}{N^2} \left( N + \sum_{\substack{n,l=1\\n\neq l}}^N E \left[ \mathrm{e}^{ikX_n} \right] E \left[ \overline{\mathrm{e}^{ikX_l}} \right] \right) \\
	&= \frac{1}{N^2} \left( N + N(N-1) |\phi(k)|^2 \right) \\
	&= |\phi(k)|^2 + \frac{1}{N} \left(1- |\phi(k)|^2 \right).
\end{align*}
Then, equation~\eqref{eq:third} can be obtained as
\begin{align*}
	& E\left[ \left | \phi(k) - \phi_N (k) \right |^2 \right ] \\
	=& E\left[ \left | \phi(k) \right |^2 - \phi(k)\overline{\phi_N(k)} - \phi_N(k)\overline{\phi(k)} + \left | \phi_N(k) \right |^2 \right] \\
	=& \left | \phi(k) \right |^2 - \phi(k) E\left[\overline{\phi_N(k)}\right]
	- \overline{\phi(k)} E\left[\phi_N(k)\right] + E\left[ \left | \phi_N (k) \right |^2 \right ] \\
	=& - \left | \phi (k) \right |^2 + E\left[ \left | \phi_N (k) \right |^2 \right ] \\
	=& \frac{1}{N} \left(1- \left |\phi(k) \right|^2 \right).
\end{align*}
Thus, the expectation of distance is
\begin{align*}
	& E\left[ \frac{\Delta k}{2\pi} \sum_{i=1}^{\frac{2\pi}{\Delta k}} \left| \hat{\phi}(k_i) - \phi_N(k_i)\right|^2 \right] \\
	=& \frac{\Delta k}{2\pi} \sum_{i=1}^{\frac{2\pi}{\Delta k}} E\left[\left| \hat{\phi}(k_i) - \phi_N(k_i)\right|^2 \right ] \\
	=& \frac{\Delta k}{2\pi} \sum_{i=1}^{\frac{2\pi}{\Delta k}} \Big( \left| \hat{\phi}(k_i) \right|^2 - \hat{\phi}(k_i) E\left[ \overline{\phi_N(k_i)} \right]  \\
	&- \overline{\hat{\phi}(k_i)} E\left[ \phi_N(k_i) \right] + E\left[ \left| \phi_N(k_i) \right|^2 \right] \Big) \\
	=& \frac{\Delta k}{2\pi} \sum_{i=1}^{\frac{2\pi}{\Delta k}} \left( \left| \phi(k_i) - \hat{\phi}(k_i)\right|^2 + \frac{1}{N} \left(1- |\phi(k_i)|^2 \right) \right) \\
	=& \frac{1}{N} \left( 1 - \frac{\Delta k}{2\pi} \sum_{i=1}^{\frac{2\pi}{\Delta k}} |\phi(k_i)|^2 \right)
	+ \frac{\Delta k}{2\pi} \sum_{i=1}^{\frac{2\pi}{\Delta k}} \left( \left| \phi(k_i) - \hat{\phi}(k_i)\right|^2 \right),
\end{align*}
which denotes that the order is $\mathcal{O}(1/N)$.
Note that Figure~\ref{distance1} (a) is the case where there is no estimation errors, where estimates $\hat{\phi}(k_i)$ is equal to the true CF $\phi(k_i)$ for all $i$.
%fitting data with alternative distributions
\section{Fitting Local Tails with Alternative One-sided Distributions}%%%%%%%%%%%%%%%%%%%
This Appendix shows the results of the best fit with two types of one-sided distribution models, the power-law and exponential distributions, for cryptocurrency data.
Before showing the results, we provide a technical description of the goodness-of-fit test.

\begin{table}
\begin{center}
\caption{Estimation results of the power-law and exponential fit with standardized cryptocurrency returns for the {\it positive} tail and its goodness-of-fit test with the $p$-value.
$\Delta t$ is the time interval, $\hat{x}_\mathrm{min}$ is the lower bound of the estimated tail, and $n_\mathrm{tail}$ represents the number of data used for the estimation (data number of the tail portion). $\hat{\alpha}$ and $\hat{\lambda}$ is the estimated power-law parameter and the estimated exponential parameter, respectively.
 Statistically significant fit $(p \geq 0.1)$ are shown in bold.
 }
   \begin{tabular}{l l p{0mm} c p{0mm} c p{0mm} c p{0mm} c} \hline
     Power law & $\Delta t$ & & $n_\mathrm{tail}$ & & $\hat{x}_\mathrm{min}$ & & $\hat{\alpha}$ & & $p$ \\ \hline \hline
     BTC & 1h & & $344$ & & $5.632$ & & $3.098$ & & $\textbf{0.598}$ \\
     & 2h & & $226$ & & $4.839$ & & $2.997$ & & $\textbf{0.637}$ \\
     ETH & 1h & & $145$ & & $7.527$ & & $3.368$ & & $\textbf{0.907}$ \\
     & 2h & & $286$ & & $4.277$ & & $2.631$ & & $\textbf{0.225}$ \\
     XRP & 1h & &$169$ & & $9.176$ & & $2.876$ & & $\textbf{0.204}$ \\
     & 2h & & $133$ & & $7.403$ & & $2.662$ & & $\textbf{0.989}$ \\
     LTC & 1h & & $194$ & & $6.939$ & & $3.221$ & & $\textbf{0.999}$ \\
     & 2h & & $103$ & & $7.272$ & & $3.371$ & & $\textbf{0.654}$ \\
     XMR & 1h & & $53$ & & $8.978$ & & $3.618$ & & $\textbf{0.980}$ \\
     & 2h & & $148$ & & $5.269$ & & $3.139$ & & $\textbf{0.805}$ \\ \hline
     Exponential & $\Delta t$ & & $n_\mathrm{tail}$ & & $\hat{x}_\mathrm{min}$ & & $\hat{\lambda}$ & & $p$ \\ \hline \hline
     BTC & 1h & & $1121$ & & $3.106$ & & $0.442$ & & $0.000$ \\
     & 2h & & $539$ & & $3.046$ & & $0.472$ & & $0.012$ \\
     ETH & 1h & & $245$ & & $5.916$ & & $0.328$ & & $\textbf{0.376}$ \\
     & 2h & & $4342$ & & $0.003$ & & $0.673$ & & $0.000$ \\
     XRP & 1h & & $346$ & & $6.220$ & & $0.234$ & & $0.016$ \\
     & 2h & & $266$ & & $4.985$ & & $0.260$ & & $0.000$ \\
     LTC & 1h & & $497$ & & $4.553$ & & $0.365$ & & $0.004$ \\
     & 2h & & $163$ & & $5.758$ & & $0.334$ & & $\textbf{0.756}$ \\
     XMR & 1h & & $8459$ & & $0.0002$ & & $0.696$ & & $0.000$ \\
     & 2h & & $4350$ & & $0.0001$ & & $0.707$ & & $0.000$ \\
     \hline
  \end{tabular}
  \label{tb:powerexp_basic_right}
  \end{center}
\end{table}
\begin{table}
 \begin{center}
 \caption{Estimation results of the power-law and exponential fit with standardized cryptocurrency returns for the {\it negative} tail (in absolute values) and its goodness-of-fit test with the $p$-value.
 Statistically significant fit $(p \geq 0.1)$ are shown in bold.
 }
   \begin{tabular}{l l p{0mm} c p{0mm} c p{0mm} c p{0mm} c} \hline
     Power law & $\Delta t$ & & $n_\mathrm{tail}$ & & $\hat{x}_\mathrm{min}$ & & $\hat{\alpha}$ & & $p$ \\ \hline \hline
     BTC & 1h & & $229$ & & $7.279$ & & $3.156$ & & $\textbf{0.943}$ \\
     & 2h & & $341$ & & $4.258$ & & $2.567$ & & $\textbf{0.301}$ \\
     ETH & 1h & & $345$ & & $5.535$ & & $2.950$ & & $0.044$ \\
     & 2h & & $414$ & & $3.612$ & & $2.353$ & & $0.001$ \\
     XRP & 1h & & $329$ & & $5.833$ & & $2.571$ & & $\textbf{0.827}$ \\
     & 2h & & $221$ & & $5.043$ & & $2.547$ & & $\textbf{0.234}$ \\
     LTC & 1h & & $128$ & & $7.882$ & & $3.778$ & & $\textbf{0.914}$ \\
     & 2h & & $149$ & & $5.743$ & & $3.168$ & & $\textbf{0.420}$ \\
     XMR & 1h & & $379$ & & $4.620$ & & $2.983$ & & $\textbf{0.135}$ \\
     & 2h & & $61$ & & $7.557$ & & $3.993$ & & $\textbf{0.902}$ \\ \hline
     Exponential & $\Delta t$ & & $n_\mathrm{tail}$ & & $\hat{x}_\mathrm{min}$ & & $\hat{\lambda}$ & & $p$ \\ \hline \hline
     BTC & 1h & & $1178$ & & $3.140$ & & $0.380$ & & $0.000$ \\
     & 2h & & $891$ & & $2.203$ & & $0.444$ & & $0.002$ \\
     ETH & 1h & & $1753$ & & $2.202$ & & $0.471$ & & $0.000$ \\
     & 2h & & $256$ & & $4.525$ & & $0.383$ & & $\textbf{0.755}$ \\
     XRP & 1h & & $71$ & & $10.46$ & & $0.174$ & & $\textbf{0.807}$ \\
     & 2h & & $80$ & & $7.560$ & & $0.245$ & & $\textbf{0.597}$ \\
     LTC & 1h & & $700$ & & $3.806$ & & $0.420$ & & $0.005$ \\
     & 2h & & $459$ & & $3.331$ & & $0.449$ & & $0.031$ \\
     XMR & 1h & & $8827$ & & $0.009$ & & $0.731$ & & $0.000$ \\
     & 2h & & $4278$ & & $0.018$ & & $0.702$ & & $0.000$ \\
     \hline
  \end{tabular}
  \label{tb:powerexp_basic_left}
  \end{center}
\end{table}

Concerning the case of the power-law distributions, as mentioned in subsection 2.2, the method of fitting data with power-laws relies on the combination of KS statistics and the maximum likelihood estimators (Hill estimator) suggested by Clauset et al.~\cite{Clauset2009}.
They also propose how to test the goodness-of-fit to see whether the hypothetical model is plausible for fitting the data.
The idea is based on the resampling method with the procedures given as follows.
After fitting the data with the power-law model, we generate synthetic datasets that follow a power law with the parameter estimate $\hat{\alpha}$ and the lower bound $\hat{x}_\mathrm{min}$.
The same method of fitting with power laws is applied again to these datasets to obtain synthetic distances between the generated CDF and the newly estimated CDF associated with the minimum KS statistic.
A sufficiently large number ($L=1000$) of synthetic datasets are generated, and each synthetic distance $D_i\ (i=1,\ldots,L)$ is then compared with the empirical distance $D$.
Finally, the $p$-value for the null hypothesis that the data follows the estimated model is calculated by using the number of times that satisfy $D_i \geq D\ (i=1,\ldots,L)$.
The confidence level is set at 90\%, in other words, if $p \geq 0.1$, we can say that the model shows a plausible fit.
Regarding the case of the exponential distributions, we conduct similar procedures as the case of the power-law distributions.

Table~\ref{tb:powerexp_basic_right} shows the results of fitting the data of standardized returns, presented in subsection 3.2, with power-law and exponential models for the positive tail, and Table~\ref{tb:powerexp_basic_left} shows the results for the negative tail.
The analysis generally confirms that the tail portion of cryptocurrency returns plausibly fit with a power law exponent with $\alpha$ approximately ranging from 2.5 to 3.5, which is slightly higher than the finding in the previous study for the Bitcoin (BTC)~\cite{Begusic2018}.
On the other hand, the exponential model is not appropriate for many cases. However, since the exponential model sometimes shows a plausible fit with empirical data, we consider it as an critical model.
Thus, we discuss the model-fit for the tail portion of returns by comparing the stable distribution with the exponential distribution in subsection 3.4.

\end{document}